\documentclass[aps,a4paper,nobibnotes,nofootinbib,amsfonts,amssymb,amsmath,eqsecnum,tightenlines,11pt]{revtex4}
\usepackage[small,center,compact,outermarks]{titlesec}
\usepackage{amssymb,amsmath}
\usepackage{hyperref}
\usepackage{graphicx}
\usepackage{placeins}
\usepackage{float}
\titlespacing{\section}{1 em}{1 em}{1 em}
\titlespacing{\subsection}{1 em}{1em}{1 em}
\titlespacing{\subsubsection}{1 em}{1em}{1 em}
\hypersetup{backref,colorlinks=true,linkcolor=blue,citecolor=red,linktocpage=true,breaklinks}

\DeclareMathAlphabet{\mathbfit}{OML}{cmm}{b}{it}

\newcommand{\sech}{\operatorname{sech}}

\begin{document}
\title{\textcolor{blue}{Nonlinear localized modes in $\mathcal{PT}$-symmetric optical media with competing gain and loss} \vspace{.35 cm}}
\author{\bf Bikashkali Midya$^1$}
\email{bikash.midya@gmail.com}
\author{\bf Rajkumar Roychoudhury$^2$}
\email{rroychoudhury123@gmail.com}
\affiliation{$^1$Physics and Applied Mathematics Unit, Indian Statistical Institute, Kolkata 700108, India. \\
$^2$Advanced center for nonlinear and complex phenomena, Kolkata 700075, India. \vspace{.5 cm}}

\begin{abstract}
 { The existence and stability of the nonlinear spatial localized modes are investigated in parity-time symmetric optical media 
 characterized by a generic complex hyperbolic refractive index distribution with competing gain and loss profile.
 The exact analytical expression of the localized modes 
 are found for all values of the competing parameter and in the presence of both the self-focusing and self-defocusing Kerr nonlinearity. The 
 effect of competing gain/loss profile on the stability structure of these localized modes
 are discussed with the help of linear stability analysis followed by the direct numerical 
 simulation of the governing equation. The spatial localized modes in two-dimensional geometry as well as the transverse power-flow 
density associated with these localized modes are also examined.}
\end{abstract}
\pacs{}
\keywords{}
 \maketitle
 \vspace{.25 cm}
 
 \section{Introduction }
 
 In the last few years considerable research has been done in the light propagation in parity-time ($\mathcal{PT}$) 
symmetric optical media \cite{Ru+10,Ko10,Gu+09,Re+12,Ma08,Mu+08a,Be08,Lo09,Lo10a,Ka+10,Ga07,We+10,LCV11,Be+10,KGM08,LK+12,LZ+13,Sc10,Sc11}. These $\mathcal{PT}$-symmetric optical structures deliberately exploit the 
quantum mechanical notion of parity $(\mathcal{P})$ and time reversal $(\mathcal{T})$ symmetry of a non-Hermitian Hamiltonian.
Non-Hermitian Hamiltonian having unbroken $\mathcal{PT}$-symmetry possesses entirely real and positive energy eigenvalues
and may constitute a physically viable system without violating any of the 
axioms of quantum mechanics \cite{BB98,BB03}. In fact, some $\mathcal{PT}$-symmetric complex
Hamiltonians has a threshold above which the corresponding energy eigenvalues become complex and the system
undergoes a phase transition because of spontaneous breaking of $\mathcal{PT}$-symmetry \cite{Be07}. For the realization
of the mathematical concept of $\mathcal{PT}$-symmetry in real settings, one can
make use of the quantum-optical analogy that the linear propagation equation
derived for optical beams in the paraxial approximation
is formally equivalent to the quantum mechanical Schr\"odinger equation. In optics $\mathcal{PT}$-symmetric structures have been constructed by judiciously incorporating balanced 
gain and loss profile in the refractive index distribution. For such settings it is necessary that the complex refractive index obeys 
the condition $n(r) = n^*(-r)$ i.e., the real and imaginary parts of the refractive index profiles must be
symmetric and anti-symmetric functions of the transverse coordinate $x$, respectively. Optical systems endowed
with this symmetry are known to exhibit previously
unattainable distinctive light propagation features such as power oscillation, band merging, double refraction, abrupt phase transitions, 
non reciprocity \cite{Ma08,Ma+10,MRR10}, unidirectional invisibility \cite{Li+11}, mode selection in $\mathcal{PT}$-symmetric lasers \cite{ML12},
as well as existence of coherent lasing absorbing
modes \cite{Lo10b}.\\

  A natural extension of the analysis of the linear $\mathcal{PT}$-symmetric structures is the investigation of the existence and stability of the
 optical solitons in a variety of nonlinear $\mathcal{PT}$-symmetric systems. In the nonlinear domain, a class of 
both the one- and two-dimensional (1D and 2D) localized modes, of the nonlinear Schr\"odinger like equation (NLS) in
the presence of Kerr nonlinearity, are found below and above the $\mathcal{PT}$
threshold of the external potential \cite{Mu+08a,Mu+08b}. The effect of nonlinearity on the stability of the fundamental spatial solitons is 
investigated in a number of complex potentials e.g. 
complex $\mathcal{PT}$-symmetric periodic \cite{Fa+12,NGY12}, hyperbolic Scarf \cite{Sh+11,KMB12}, Gaussian \cite{Hu+11}, Bessel \cite{HH12},
and parabolic potentials \cite{ZK12}. The presence of gap solitons in competing cubic-quintic
media \cite{LZ12}, defect nonlinear modes in linear lattice \cite{WW11}, gray solitons \cite{Li+11d} and multi-pole 
solitons \cite{HL13} in Kerr nonlinear media in the presence of extended gain/loss are reported. 
The existence of dark soliton and vortices described by a harmonic trap with a rapidly
decaying $\mathcal{PT}$-symmetric imaginary component in the presence of defocusing nonlinearity is also studied \cite{Ac+12}. The $\mathcal{PT}$ dipole embeded into the Kerr nonlinear medium has been shown to admit a full family of analytical soliton solutions \cite{MMR13}.
Furthermore, the spatial and temporal solitons and their stability in the $\mathcal{PT}$-symmetric coupler with balanced gain/loss are examined in 
\cite{DM11,Al+12}.
In most of these events the complex potentials have localized (asymptotically vanishing) imaginary part which 
render the associated solitons to propagate stably 
under appropriate restriction on the gain/loss parameter. In a recent study \cite{MR13}, however, we have shown that the nonlinear modes 
in a complex Rosen-Morse potential well 
are always unstable although spontaneous breakdown of $\mathcal{PT}$-symmetry is absent in such system. 
The reason behind this is that the gain/loss profile 
for such systems is asymptotically non-vanishing and any small fluctuations of the initial field are
amplified which leads to the instability.

In this paper we have investigated the optical beam propagation in a single $\mathcal{PT}$ cell which is characterized
by the nonlinear Schr\"odinger equation with a generic complex hyperbolic potential 
(given in equation (\ref{e2})) with competing gain and loss profile.  The motivation behind the study of the present model is two fold. The first one is that the beam dynamics 
in a single $\mathcal{PT}$ potential is useful to understand the light self trapping in complex optical lattices. Most importantly, one can investigate 
the effect of 
asymptotically vanishing and non-vanishing gain and loss profiles on the stable propagation of solitons in the same framework by 
considering the special cases of the competing parameter $k$. Secondly, the class of potentials 
considered here belongs to the set of few potentials for which the NLS equation admits analytical localized solutions. We find that exact 
analytical expression of the spatial localized modes can be
obtained for all values of the competing parameter and in the presence of 
self focusing and self-defocusing Kerr nonlinearity. The linear stability 
analysis followed by the direct numerical simulation have been performed to study the stability of these localized modes. 
Emphasis has been given to the cases $k = 0$ and $k > 0$ to investigate the effect of competing parameter on the stability structure of the solitons supported by the asymptotically
vanishing and non-vanishing gain/loss profiles, respectively. The spatial localized modes in two-dimensions as well as the transverse power-flow 
density associated with these localized modes are also discussed.

\section{Localized Modes in 1D $\mathcal{PT}$-symmetric optical media with competing gain and loss}

\subsection{Mathematical Formulation}

 Consider the optical beam propagation along longitudinal $z$ axis in a Kerr nonlinear media with a transverse complex $\mathcal{PT}$-symmetric 
 refractive index distribution.
 The dynamics of the beam is governed by the $(1+1)$-dimensional NLS equation with external potential \cite{Ga07,Mu+08a}
 \begin{equation}
i \frac{\partial \Psi}{\partial z} + \frac{\partial^2 \Psi}{\partial x^2} + \left[ V(x)+ i W(x)\right] \Psi + \sigma |\Psi|^2 \Psi =0, \label{e1}
\end{equation}
where $\Psi(x,z)$ is proportional to the electric field envelop, and $\sigma= +1$ corresponds to a self-focusing nonlinearity and $\sigma = -1$ to a defocusing one.
$V(x)$ and $W(x)$ are the real and imaginary parts of the complex potential 
which satisfies the $\mathcal{PT}$-symmetry i.e. $V(-x) = V(x)$ and $W(-x) = -W(x)$.
Furthermore, in the context of optics, $V(x)$ determines the bending and slowing down of light, and
$W(x)$ is responsible for either amplification or absorption of light within an optical
material.

Here we investigate optical beam propagation in a single $\mathcal{PT}$ cell which is characterized by the following class of
complex potential with competing gain and loss profile
\begin{equation}
\begin{array}{ll}
V(x) = V_0 \sech^2 x + V_1 \sech^{2k} x,\\
W(x) =  W_0 \sech^k x \tanh x,
\end{array}\label{e2}
\end{equation}
where $V_0$, $V_1$ $\in \mathbb{R}$ are the amplitude of the real part of the potential, $W_0 \in \mathbb{R}$ is 
proportional to the strength of the imaginary part and $k \in \mathbb{R}^+ \cup \{0\}$ is the competing parameter.
Each value of $k$ gives rise to a new potential whose functional forms are different. Specifically,
the nature of the gain and loss profile $W(x)$ depends much on $k$: $W(x)$ becomes asymptotically non vanishing
for $k =0$, whereas non zero values of $k$ make $W(x)$ to localize (asymptotically vanishing). 
The stationary solution of the nonlinear equation (\ref{e1}) can be obtained by assuming $\Psi(x,z) = \phi(x) e^{i \beta z}$,
where the complex valued function $\phi(x)$ describes, in general, the soliton and $\beta$ is the real propagation constant. 
Substitution of this expression of $\Psi(x,z)$ and the potential (\ref{e2}) into
Eq. (\ref{e1}) yield an ordinary nonlinear differential equation of $\phi(x)$ :
\begin{equation}
\frac{d^2 \phi}{d x^2} + \left[V_0 \sech^2 x + V_1 \sech^{2k} x  + i W_0 \sech^k x \tanh x \right] \phi + \sigma |\phi|^2 \phi = \beta \phi, \label{e3}
\end{equation}
which can be solved numerically by applying the boundary conditions $\phi_{r,i}(x) \rightarrow 0$ for sufficiently large $x$. 
The subscripts $r$ and $i$ denote `real' and `imaginary' respectively.
However, in the following we have shown that the equation (\ref{e3}) admits analytical 
localized solution for both the self-focusing and 
self-defocusing nonlinearities. 

\subsection{Analytical solution and their linear stability}
To obtain the analytical expression of the localized modes to equation (\ref{e3}), we assume
\begin{equation}
\phi(x) = \phi_0 \sech x ~e^{i \int \gamma(x) dx},
\end{equation}
where two unknown quantities $\phi_0$ and $\gamma(x)$ will be determined shortly. Substituting the above expression of $\phi(x)$ in  equation (\ref{e3}) and then separating out the real and imaginary parts we have
\begin{equation}
\begin{array}{ll}
\gamma'(x) - 2 \gamma(x) \tanh(x) + W_0 \sech^k x \tanh x =0\\
(\sigma \phi_0^2 + V_0 - 2) \sech^2 x + V_1 \sech^{2 k} x -\gamma(x)^2 = \beta - 1.
\end{array}
\end{equation}
Now, one can easily solve the above two equations to obtain 
\begin{equation}
\gamma(x) = \frac{W_0}{k+2} \sech^k x, ~~~ \phi_0 = \sqrt{\frac{2-V_0}{\sigma}},~~ \beta =1,
\end{equation}
 and hence the analytical localized solution of the equation (\ref{e3}) is reduced to
\begin{equation}
\displaystyle \phi(x) =   \sqrt{\frac{2 - V_0}{\sigma}} ~\sech x ~ e^{i \frac{W_0}{k+2} \sinh x ~ _2F_1\left(\frac{1}{2},\frac{k+1}{2},\frac{3}{2},-\sinh^2 x\right) },\label{e4}
\end{equation}
where $V_1 = \frac{W_0^2}{(k+2)^2}$ and $_2F_1(a,b,c,z)$ is the Hypergeometric function. Similar approach can be applied to find other localized solutions of the equation (\ref{e3}). Specifically, for $k \ne 0$, the equation (\ref{e3}) is found to admit another analytical solution of the form $\tilde{\phi}(x) = \phi_0 \sech^k x ~\exp\left[\frac{i W_0}{3 k} \sinh x ~ _2F_1\left(\frac{1}{2}, \frac{k+1}{2}, \frac{3}{2}, -\sinh^2 x\right)\right]$, where $V_0 = k (k+1), V_1 = \frac{W_0^2}{9 k^2} - \sigma \phi_0^2$ and $\beta = k^2$. However, the first solution given in equation (\ref{e4}) is valid for all nonnegative real values of 
$k$ including $k=0$. Therefore we shall consider the solution (\ref{e4}) in the remaining part of the paper to investigate the effect of both $k=0$ and $k \ne 0$ on the stable propagation of such solutions. In order to probe physical properties of these localized solutions
we shall examine following three quantities: 
the total power $P$, the transverse power (Poynting vector) $S$ across the beam and the linear stability analysis of those localized modes. In the 
present case, the total power is calculated as $P = \int_{-\infty}^{+\infty} |\phi(x)|^2 ~ dx = 2 (2-V_0)$. It is clear that the power is independent of the competing parameter $k$
 and remains positive for $V_0 <2$. The transverse power or energy flow of light field accross the beam is given by 
 $S = \frac{i}{2} (\phi \phi_x^* - \phi^* \phi_x) = \frac{W_0(2-V_0)}{k+2} \sech^{(k+2)} x$. Obviously $S$ depends on the competing parameter $k$ and 
it remains positive for positive values of $W_0$ and $V_0 <2$. This suggests that the power flow in the $\mathcal{PT}$ cell 
is unidirectional i.e. from gain towards loss region.

In order to investigate the effect of the competing gain and loss profile on the linear stability properties of the 
self-trapped nonlinear modes obtained here,
we consider small perturbation to the solution $\Psi(x,z)$, in the form \cite{KMB12,NGY12,ZKK11}
\begin{equation}
 \Psi(x,z) = \phi(x) e^{i \beta z} + \left\{[ f(x) + g(x) ] ~e^{\delta z} + [f^*(x) -g^*(x)]~ e^{\delta^* z}\right\}  e^{i \beta z}\label{e30}
\end{equation}
where superscript $*$ represents the complex conjugation, $|f|$, $|g| \ll |\phi|$ are the infinitesimal perturbation 
eigen-functions which may grow upon propagation with the perturbation growth rate $\delta$. Substituting the perturbed
solution into Eq. (\ref{e1}) and linearizing it around the stationary solution $\phi(x)$, one obtains a coupled
set of linear eigenvalue equations:
\begin{equation}
 \left( \begin{array}{cc}
0 & ~\hat{\mathcal{L}}_1  \\
 \hat{\mathcal{L}}_2 & ~0 \\
\end{array} \right)   ~~~ \left( \begin{array}{c}
f   \\
g \\
 \end{array} \right)= - i \delta
 \left( \begin{array}{c}
f \\
g \\
 \end{array} \right)\label{e12}
\end{equation}
where $\hat{\mathcal{L}}_1 = \partial_{xx} + (V+i W) + \sigma \phi^2 - \beta$ and
$\hat{\mathcal{L}}_2 = \partial_{xx} + (V+i W) + 3 \sigma \phi^2 - \beta$. This coupled equations can be 
solved numerically with the help of Fourier collocation method \cite{Ya08}. The linear stability of a soliton is determined by the nature of the spectrum
of the above eigenvalue problem (\ref{e12}). If there exist any $\delta$ with a positive real part, the perturbed solution (\ref{e30}) would grow exponentially with 
$z$ and thus corresponding solitons become linearly unstable.
On the other hand stationary solutions can be completely stable
only when all real parts of $\delta$ are equal to zero.

In the following we consider some particular values of $k$ for which the Hypergeometric function (appeared in the phase of the solution \ref{e4}) can be 
expressed in closed analytical form. For numerical computation we have also restricted ourself in the self focusing nonlinearity 
$(\sigma =1)$, and $V_0 =1$. Self defocusing case $(\sigma=-1)$ is discussed in subsequent section. For each of the particular choices of $k$, we have numerically investigated the linear stability analysis of the corresponding localized modes 
for arbitrary values of gain and loss parameter $W_0$. We find that for $k = 0$ the localized modes are unstable for all $W_0$. Whereas in 
other cases, 
considered here, stable propagation is obtained below a certain value of $W_0$. The results of our investigations are summarized below. 
These results are also confirmed by the direct numerical simulation 
of the governing equation (\ref{e1}) by considering the initial profile as $\Psi(x,z=0) = \phi(x)$.\\ 

\begin{figure}[h]
 \includegraphics[width=4.25 cm, height=4.5 cm]{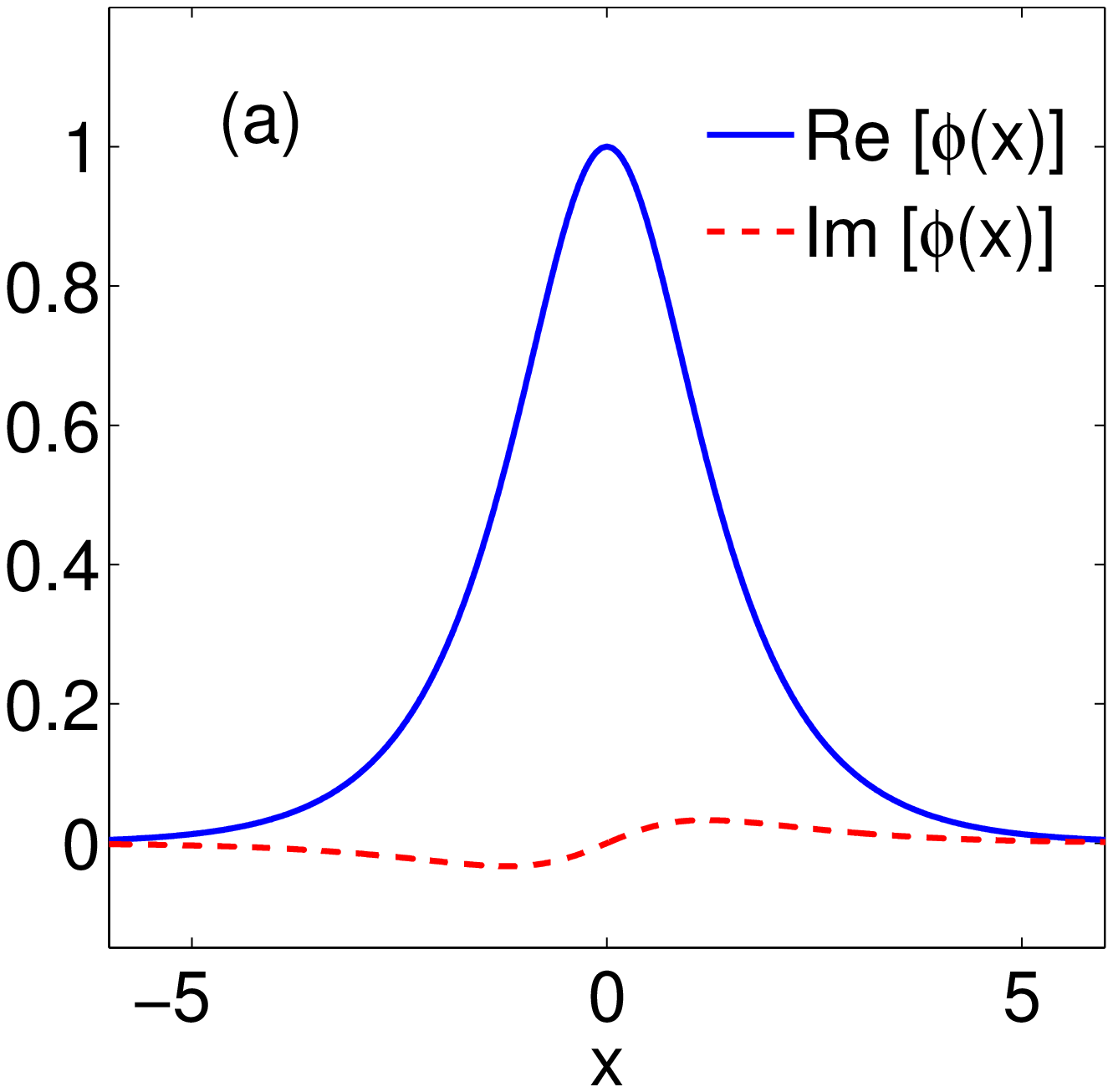} 
 \includegraphics[width=7.25 cm,height=4.5 cm]{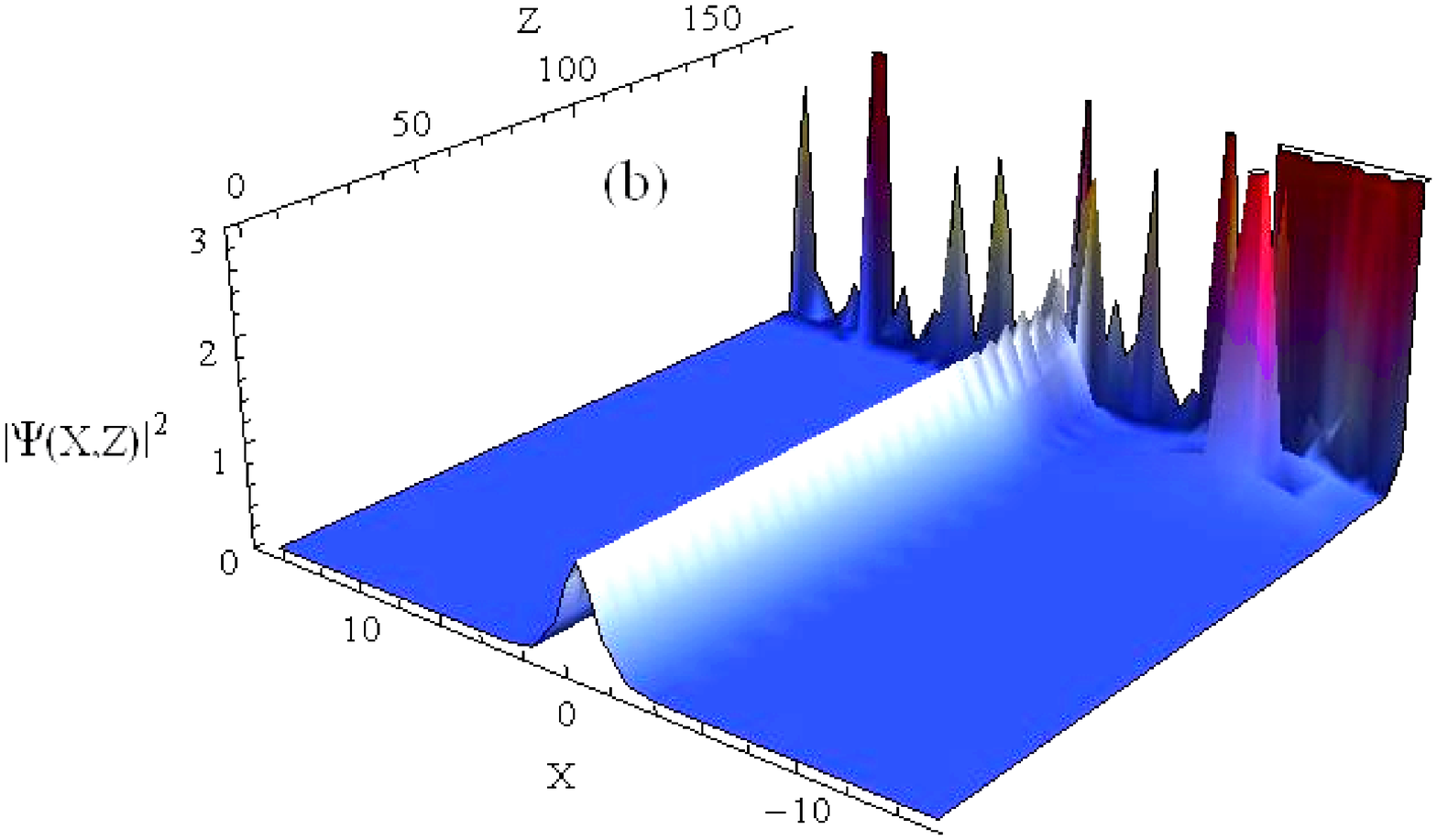} 
 \includegraphics[width=4.25 cm,height=4.5 cm]{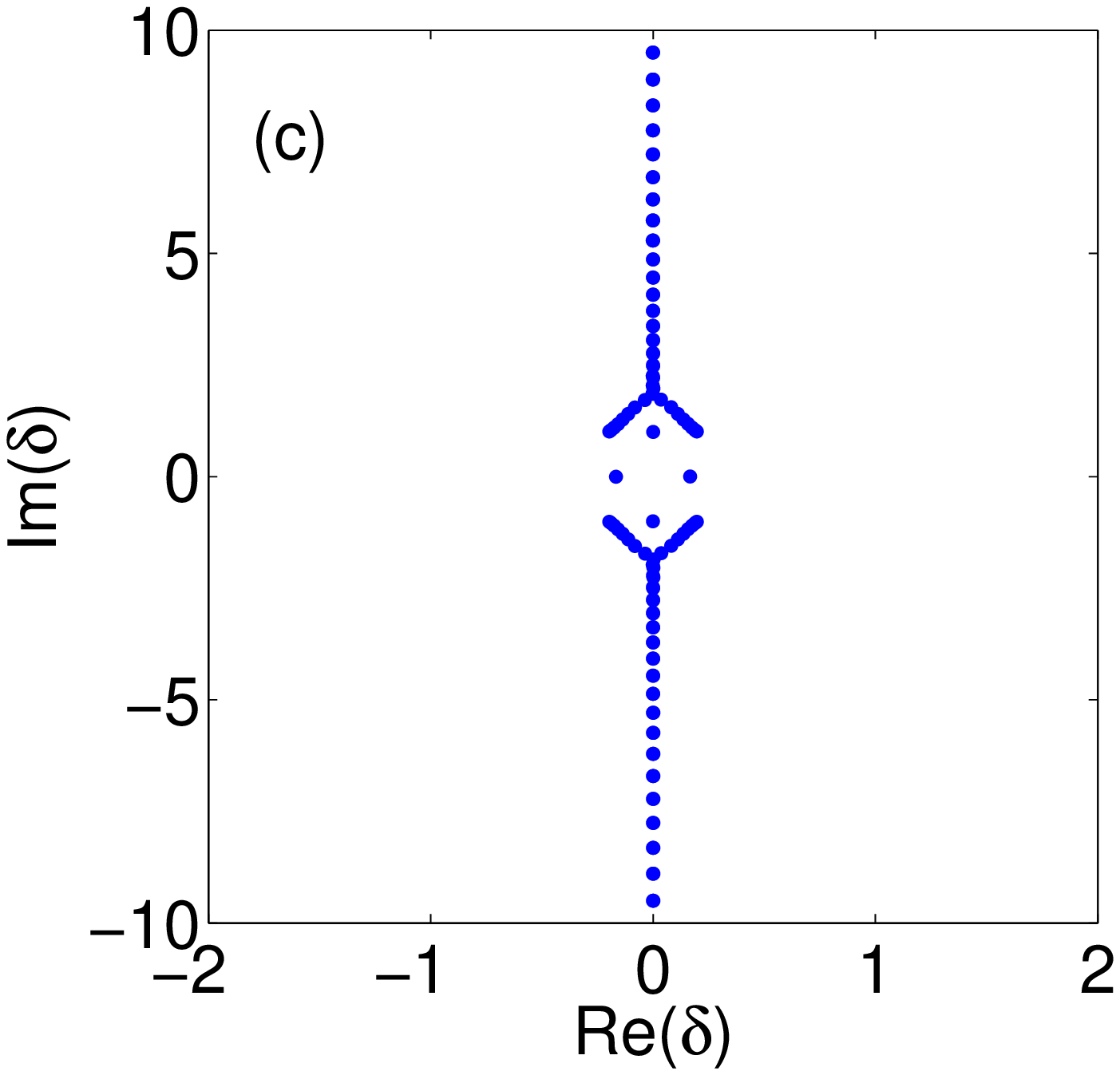}
 \caption{(Color online) Plot of (a) the real and imaginary components of the localized modes $\phi(x)$ in the complex Rosen-Morse potential 
 (b) the corresponding unstable intensity evolution $|\Psi(x,z)|^2$, and (c) numerically computed linear stability spectra. 
 We have considered here $k=0, V_0 = 1, W_0 = 0.1, \sigma=1$ and $\beta =1$. }\label{f1}
\end{figure}
\FloatBarrier

\subsubsection{ $k= 0 :$ complex Rosen-Morse potential}

For $k=0$, the potential (\ref{e2}) reduces to the complex Rosen-Morse potential $V(x) = V_0 \sech^2 x + \frac{W_0^2}{4}$ and $W(x) = W_0 \tanh x$. Clearly,
the imaginary part 
$W(x)$ of this potential is asymptotically 
non vanishing and it is known that the linear Schr\"odinger eigenvalue 
problem for this potential possesses unbroken $\mathcal{PT}$-symmetry \cite{LM09}. Nevertheless, it is found that localized modes 
$\phi(x) = \sqrt{2-V_0} ~\sech x ~e^{\frac{i W_0}{2} x}$ 
described by this complex Rosen-Morse potential are unstable and non physical for all values of $W_0$. Recently, this result has also been reported in details by us \cite{MR13}. 
In figures \ref{f1} (a),(b), we have plotted the 
real and imaginary parts 
of the nonlinear modes $\phi(x)$ and the corresponding unstable intensity evolution $|\Psi(x,z)|^2$, respectively. The numerically computed 
linear stability spectra is plotted if figure \ref{f1}(c), which clearly predicts the instability. All these figures are plotted
by considering $\sigma =1, \beta =1, V_0 = 1$ and $W_0=0.1$.

\begin{figure}[h]
 \includegraphics[width=4.25 cm, height=4.35 cm]{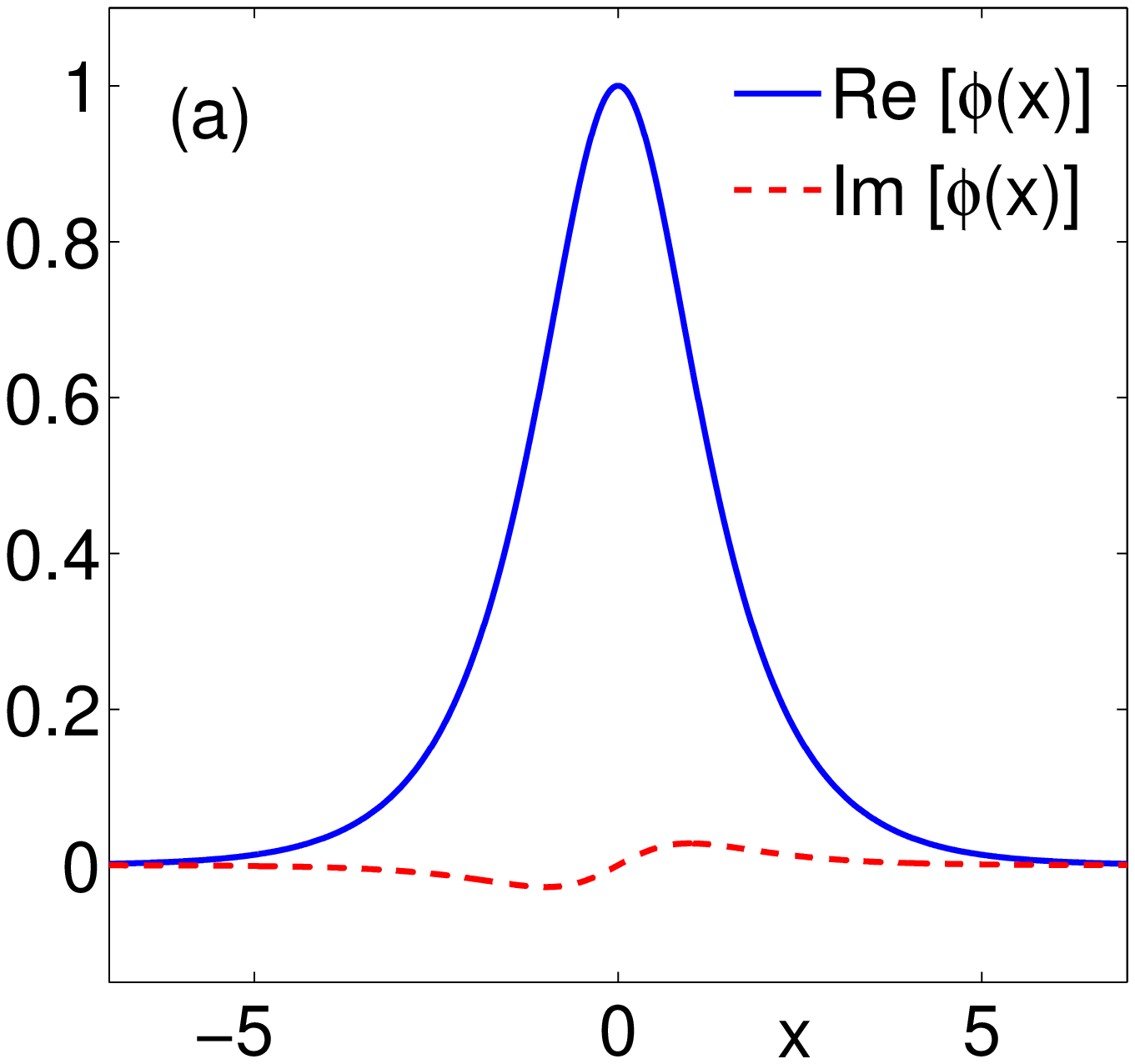} 
 \includegraphics[width=7.25 cm,height=4.35 cm]{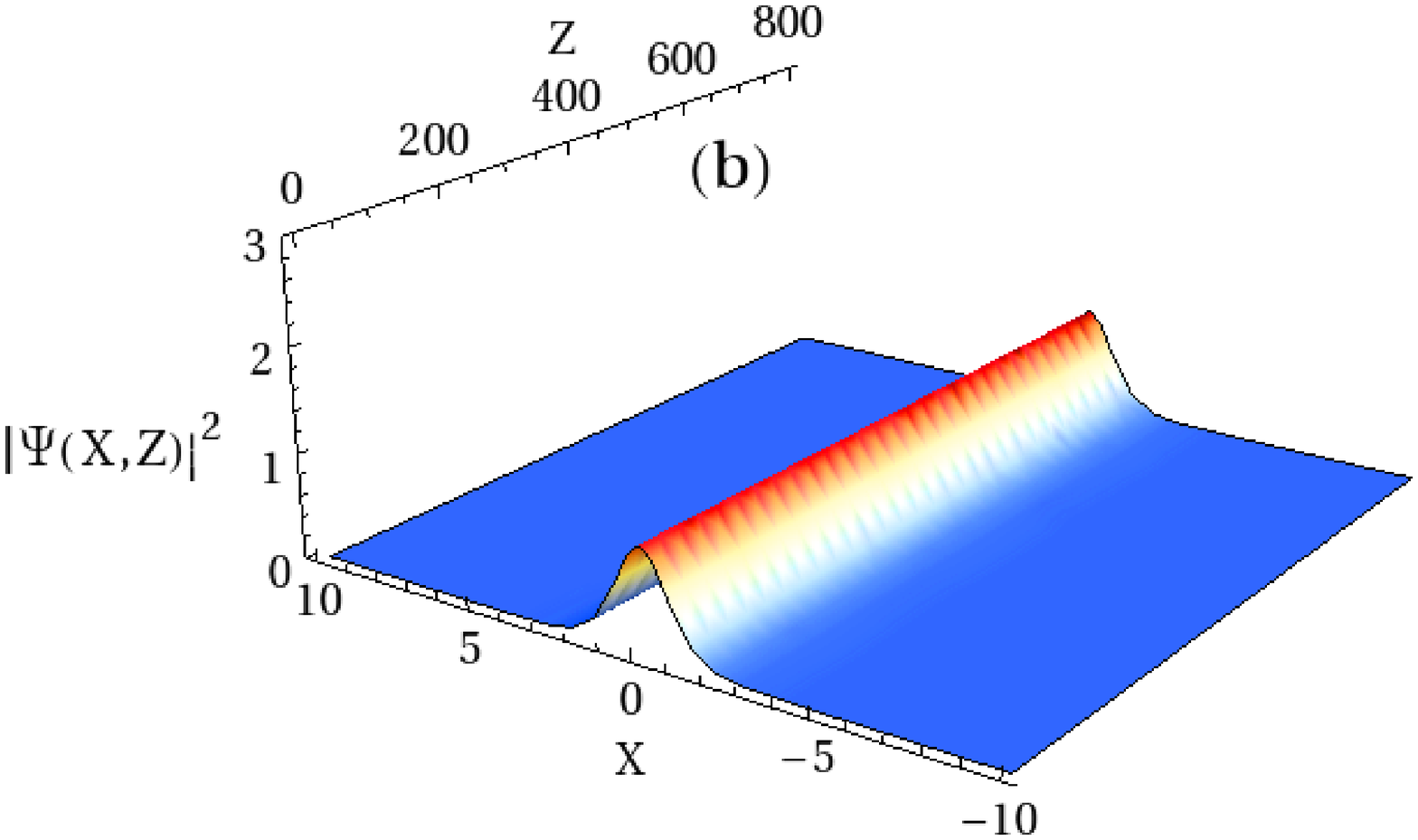} 
 \includegraphics[width=4.25 cm,height=4.35 cm]{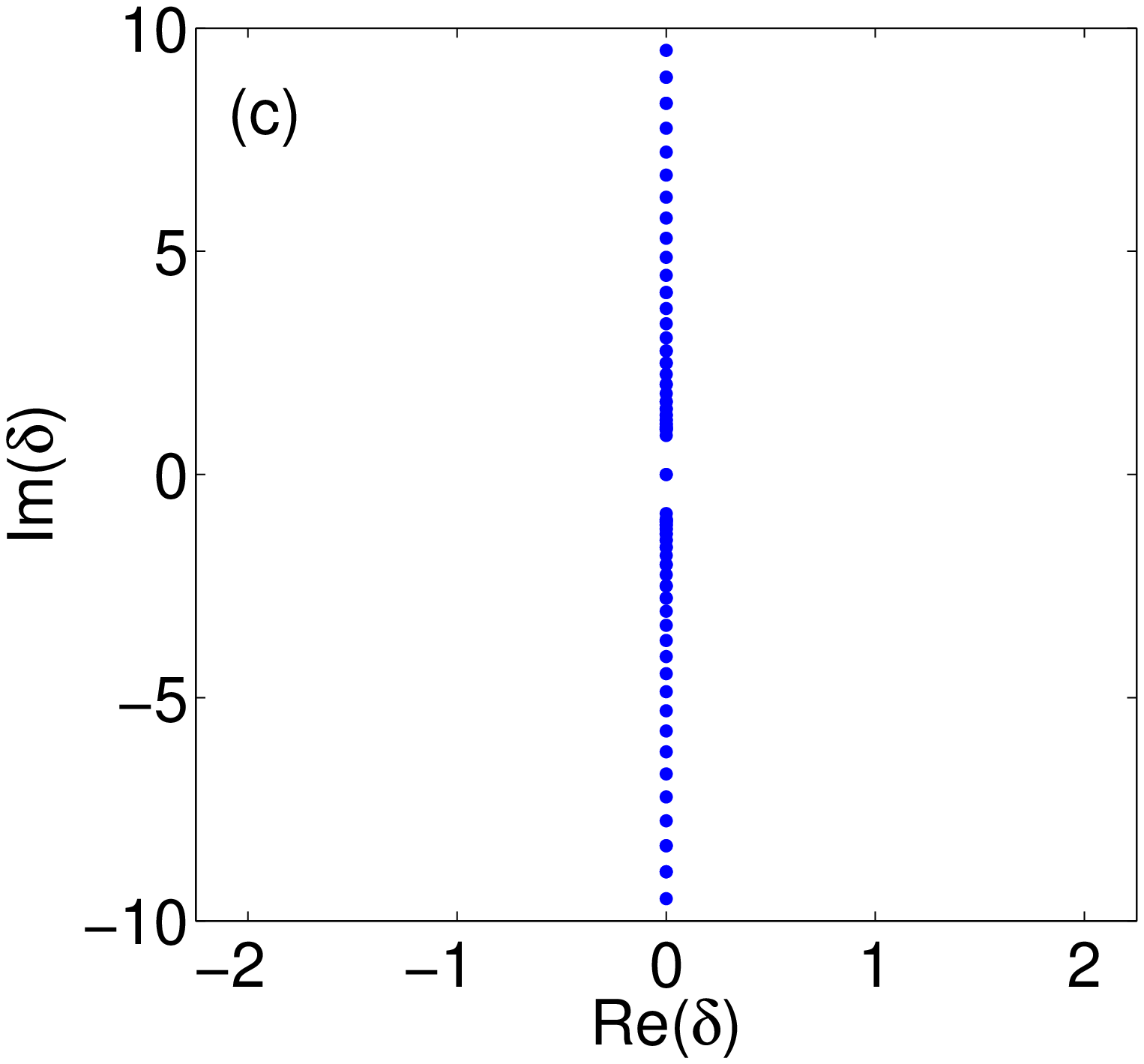}
  \includegraphics[width=4.25 cm, height=4.35 cm]{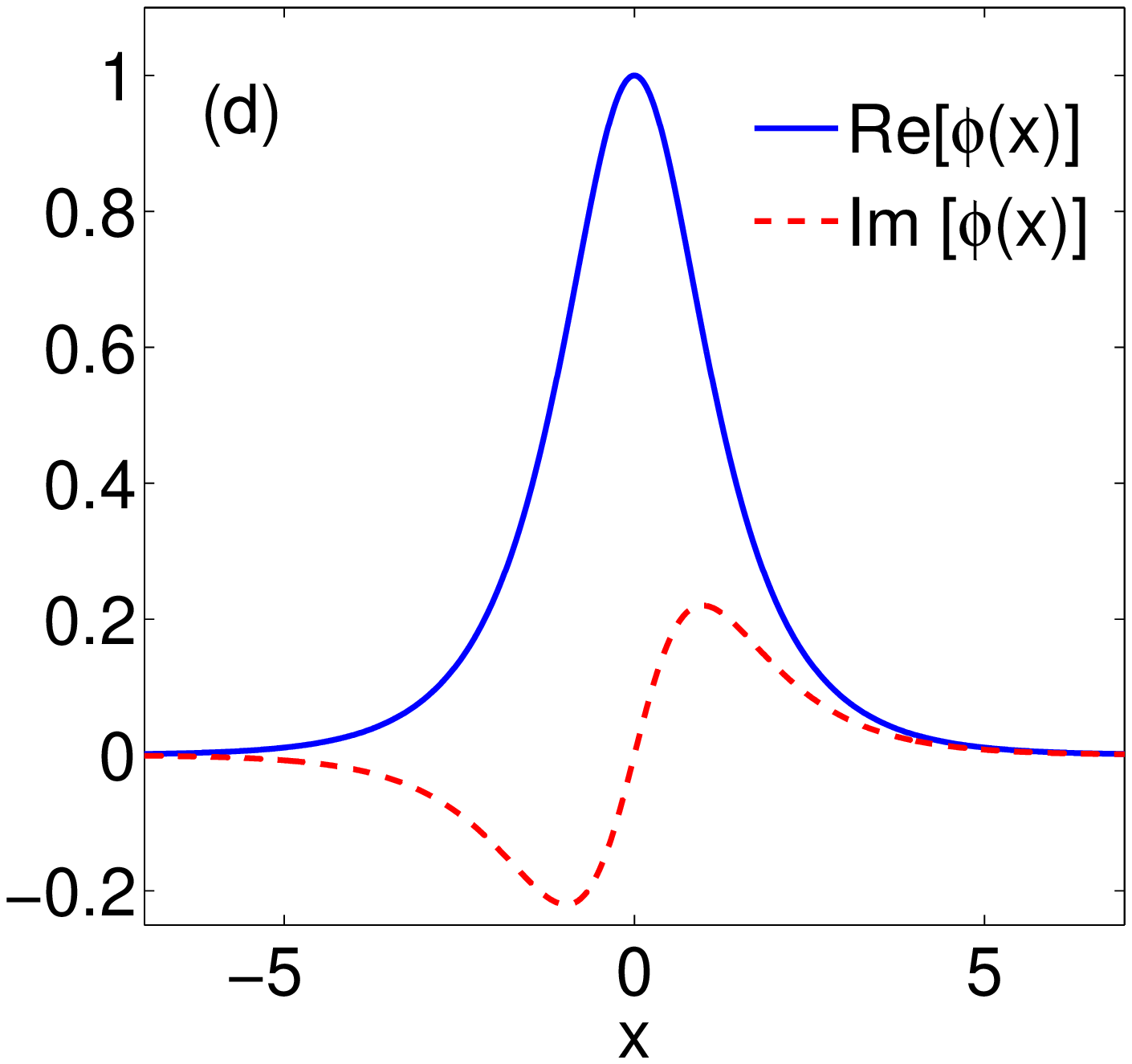} 
  \includegraphics[width=7.25 cm,height=4.35 cm]{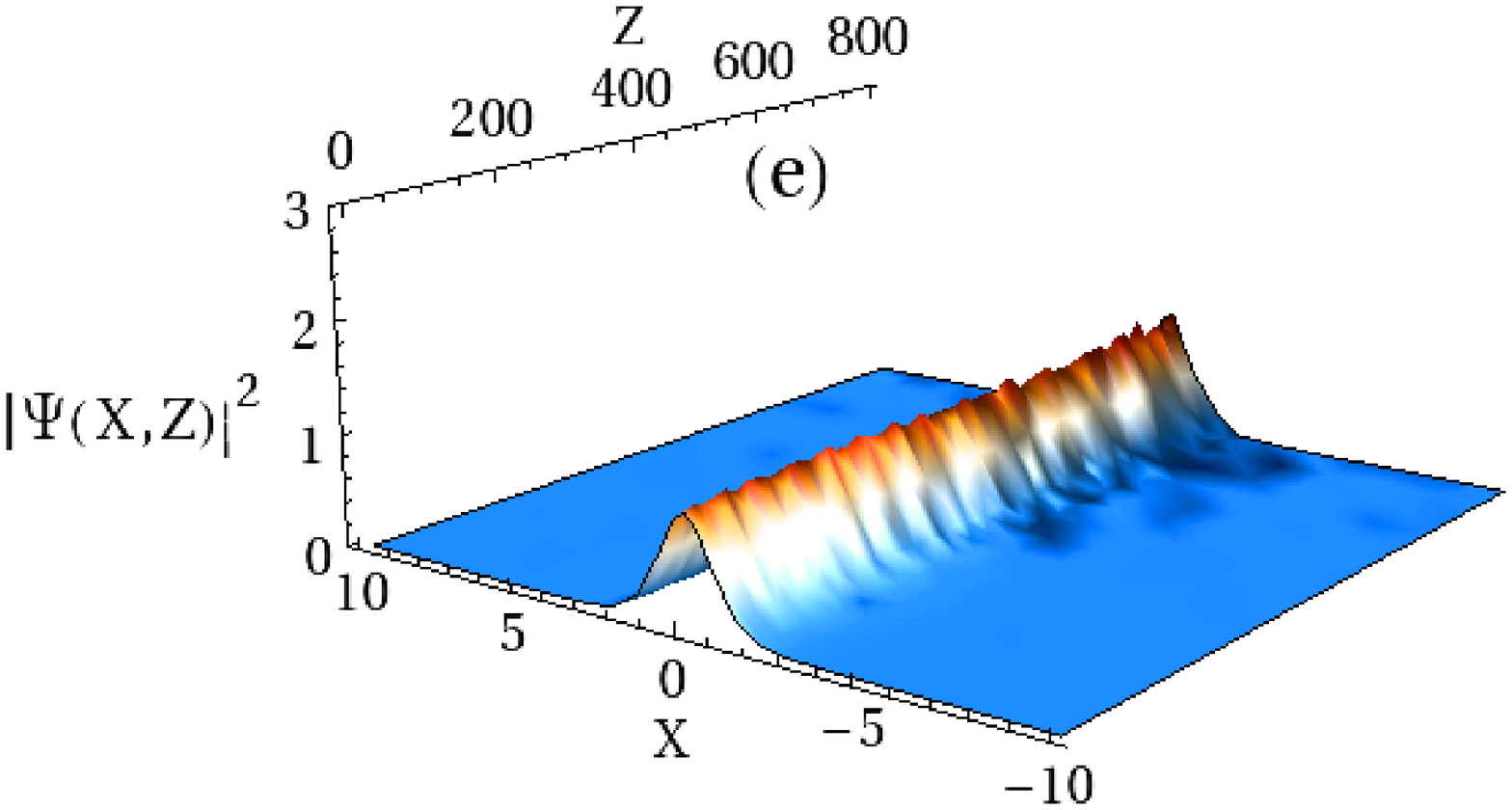} 
   \includegraphics[width=4.25 cm,height=4.35 cm]{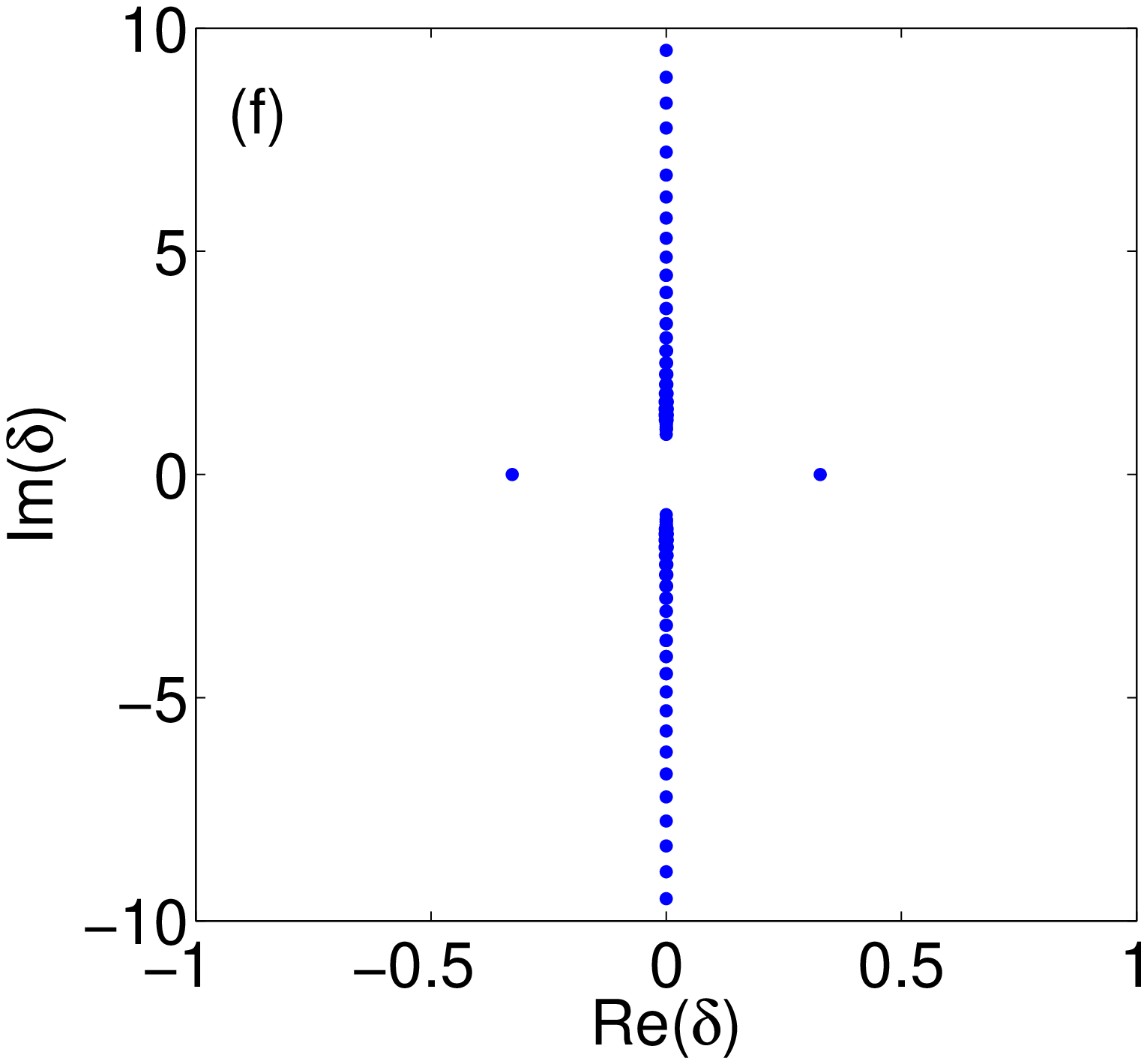}
 \caption{(Color online) (a) Real and imaginary components of the stable localized modes $\phi(x)$ described by the complex hyperbolic scarf potential (b)
 corresponding stable intensity evolution $|\Psi(x,z)|^2$, and (c) the linear stability spectra. All the figures have been plotted for 
 $k=1, V_0 =1, W_0 = 0.1, \sigma=1$ and $\beta =1$. Same have been plotted in (d), (e) (f) respectively, for the unstable localized 
 modes corresponding to 
 $W_0=.8$. All other parameters are kept unchanged.}\label{f2}
\end{figure}
\FloatBarrier

\subsubsection{$k=1 :$ complex hyperbolic Scarf potential}
In this case, we have the complex hyperbolic Scarf potential $V(x) = \left(V_0 + \frac{W_0^2}{9}\right) \sech^2 x$, $W(x) = W_0 \sech x \tanh x$. Clearly, the gain/loss profile $W(x)$
vanishes for large values of transverse co-ordinate $x$. The localized solution corresponding to this potential is given by $\phi(x) = \sqrt{2-V_0} 
~\sech x ~ e^{\frac{i W_0}{3} \tan^{-1} (\sinh x)}$. Our numerical study reveals that for $V_0 =1$ these localized modes are 
stable below the threshold $W_0 \sim 0.5$. For $W_0 > 0.5$, the corresponding localized modes become unstable. 
Same bound $W_0 =.5$ has also been obtained in reference \cite{KMB12}. In figure \ref{f2}, we have plotted 
the real and 
imaginary parts of both the stable and unstable localized solutions $\phi(x)$ below and above the threshold $W_0 = 0.5$. The 
corresponding intensity evolutions $\Psi(x,z)$ and the linear stability spectra are also shown in these figures.

\subsubsection{ $k = 3 : V(x) = V_0 \sech^2 x + \frac{W_0^2}{25} \sech^6 x, ~~ W(x) = W_0 \sech^3 x \tanh x$ }
\begin{figure}[h]
 \includegraphics[width=5.25 cm, height=4.5 cm]{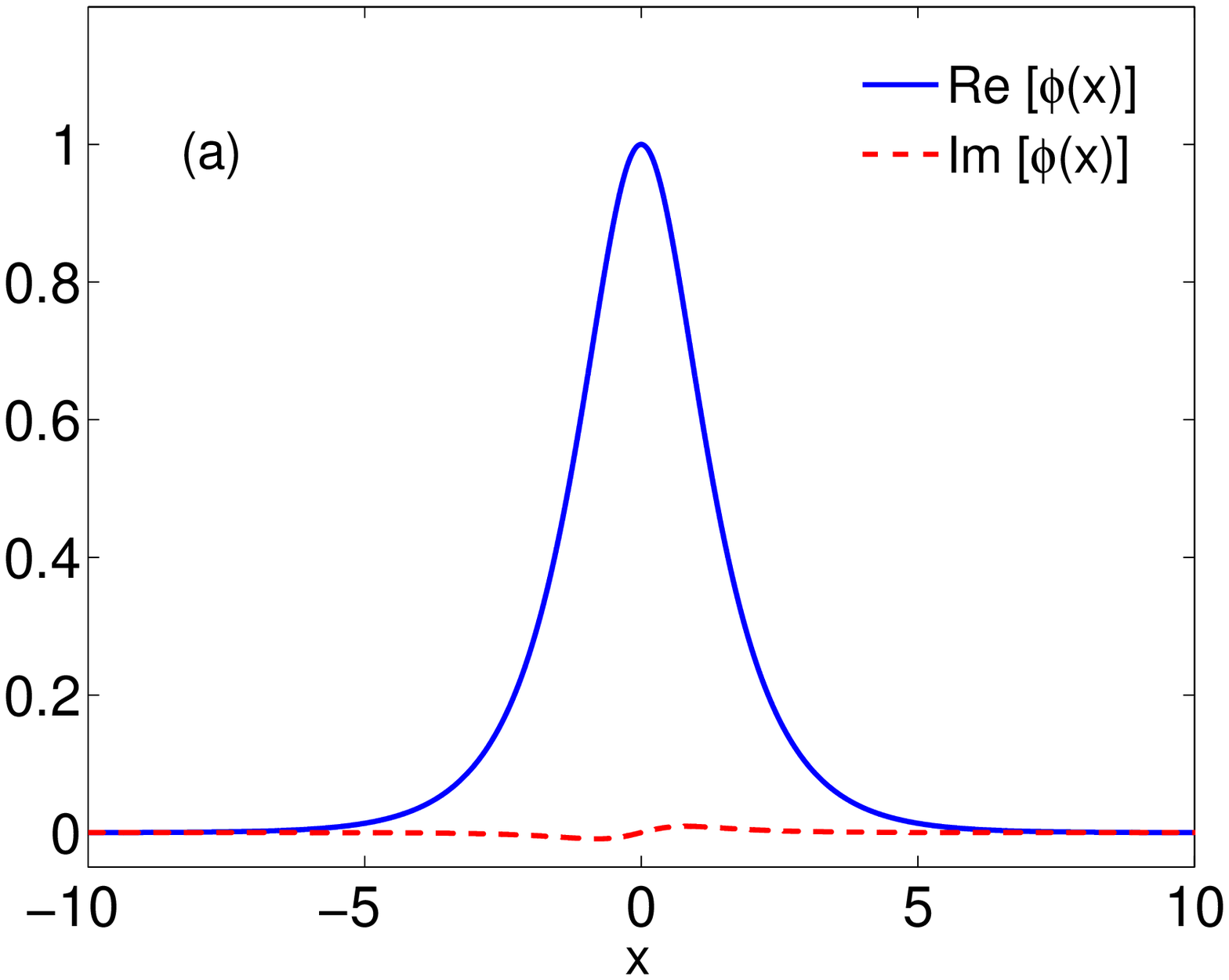} 
 \includegraphics[width=6.25 cm,height=4.5 cm]{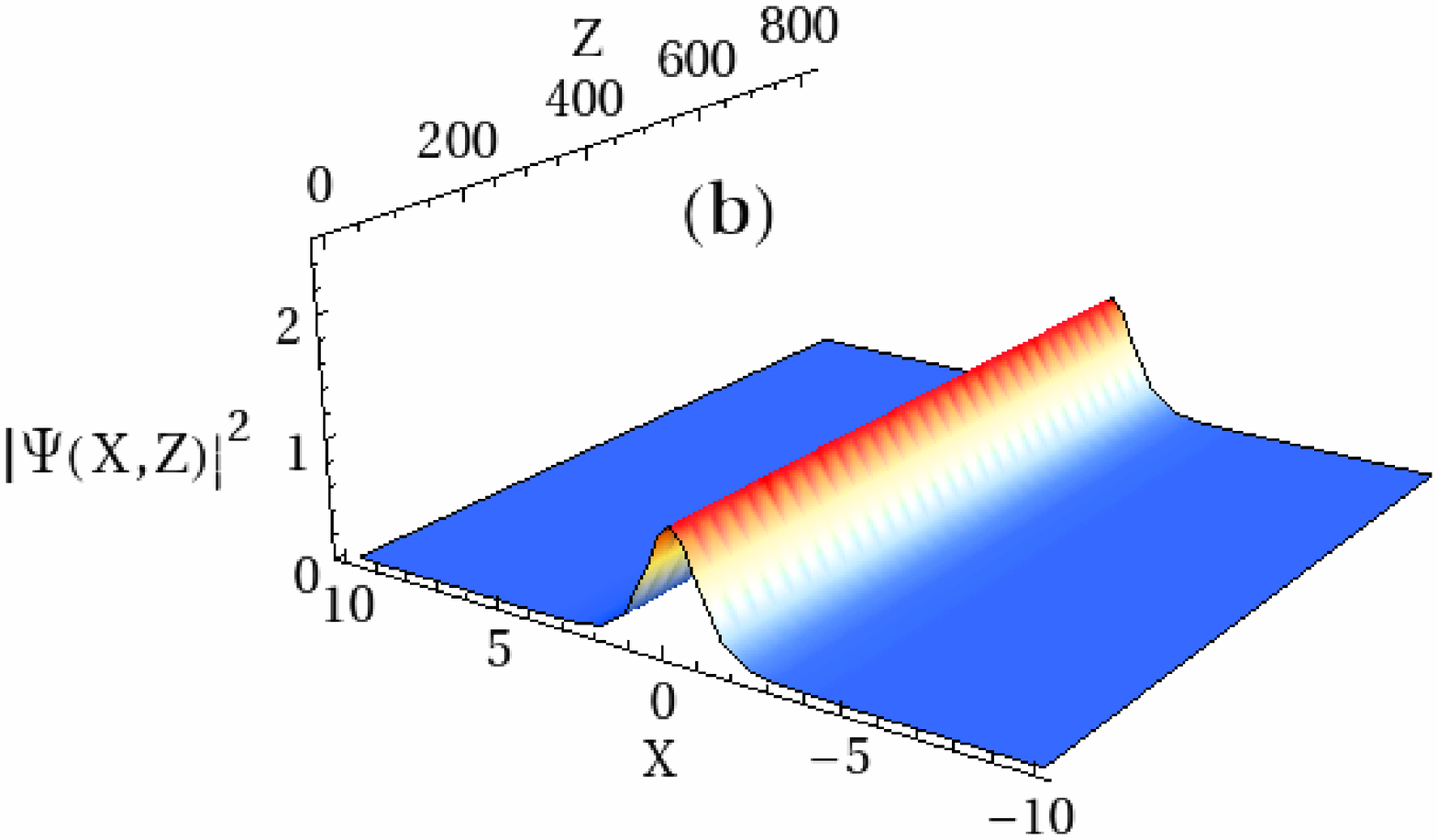} 
 \includegraphics[width=4.25 cm,height=4.5 cm]{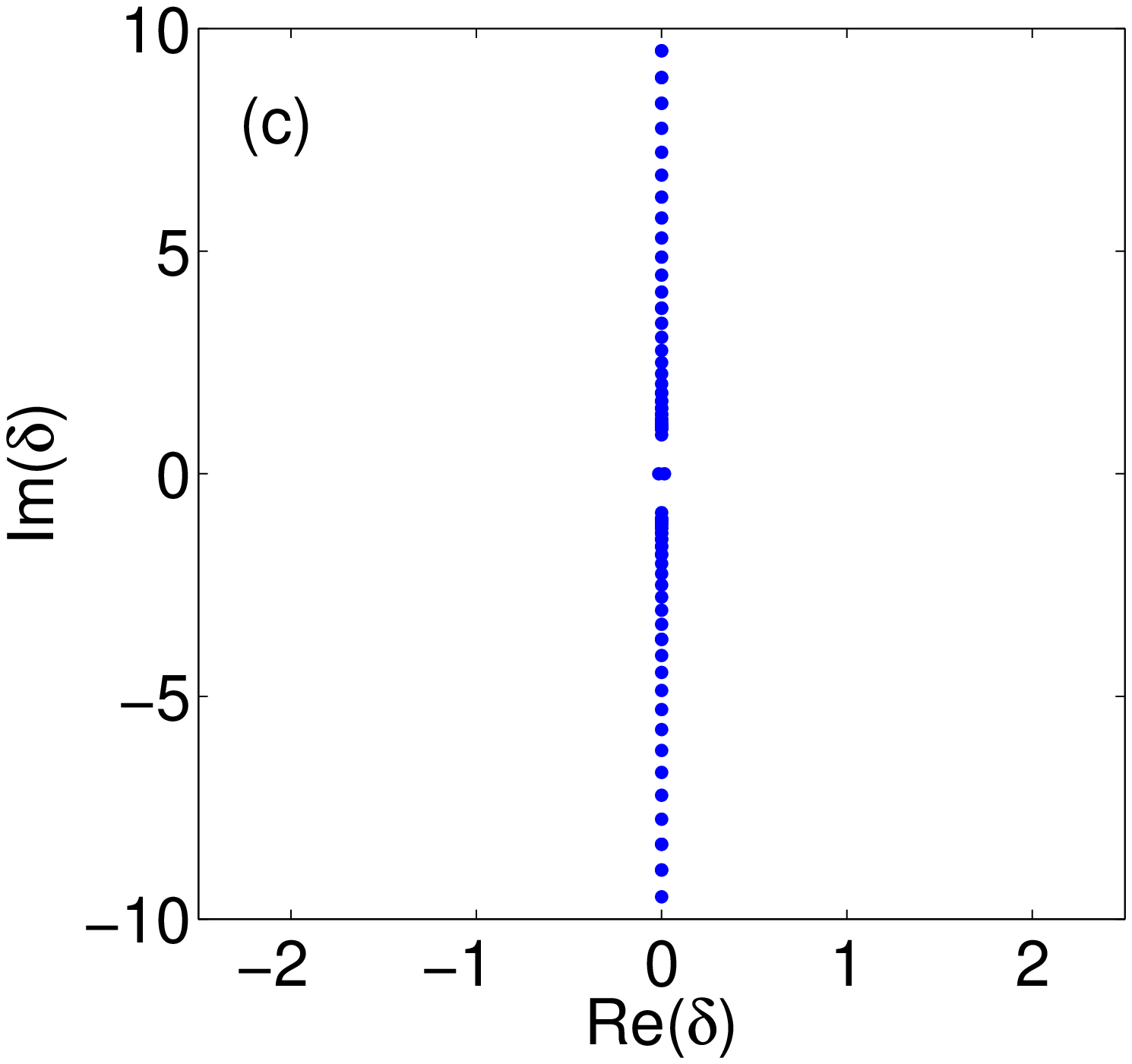}
 \includegraphics[width=5.25 cm, height=4.5 cm]{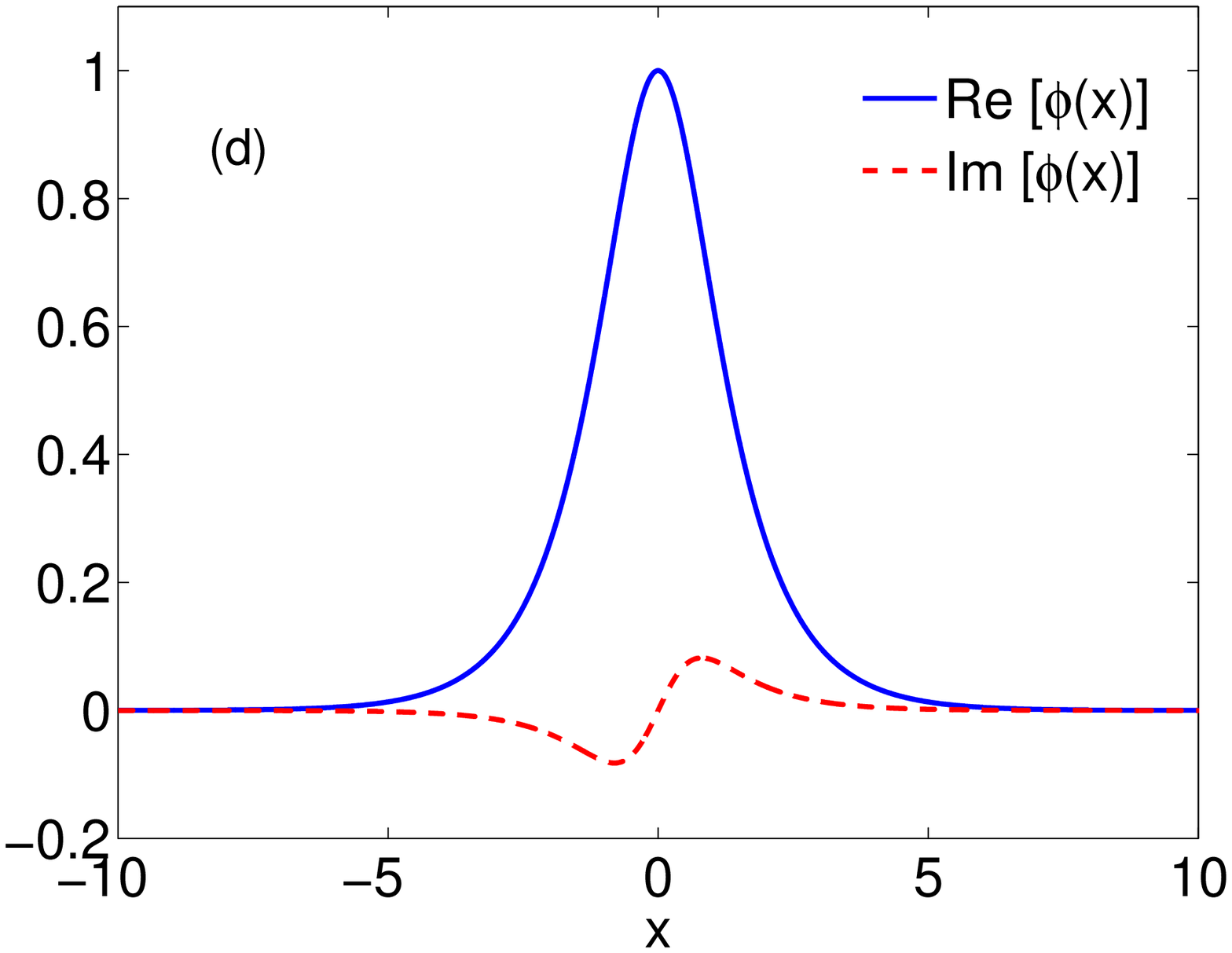} 
 \includegraphics[width=6.25 cm,height=4.5 cm]{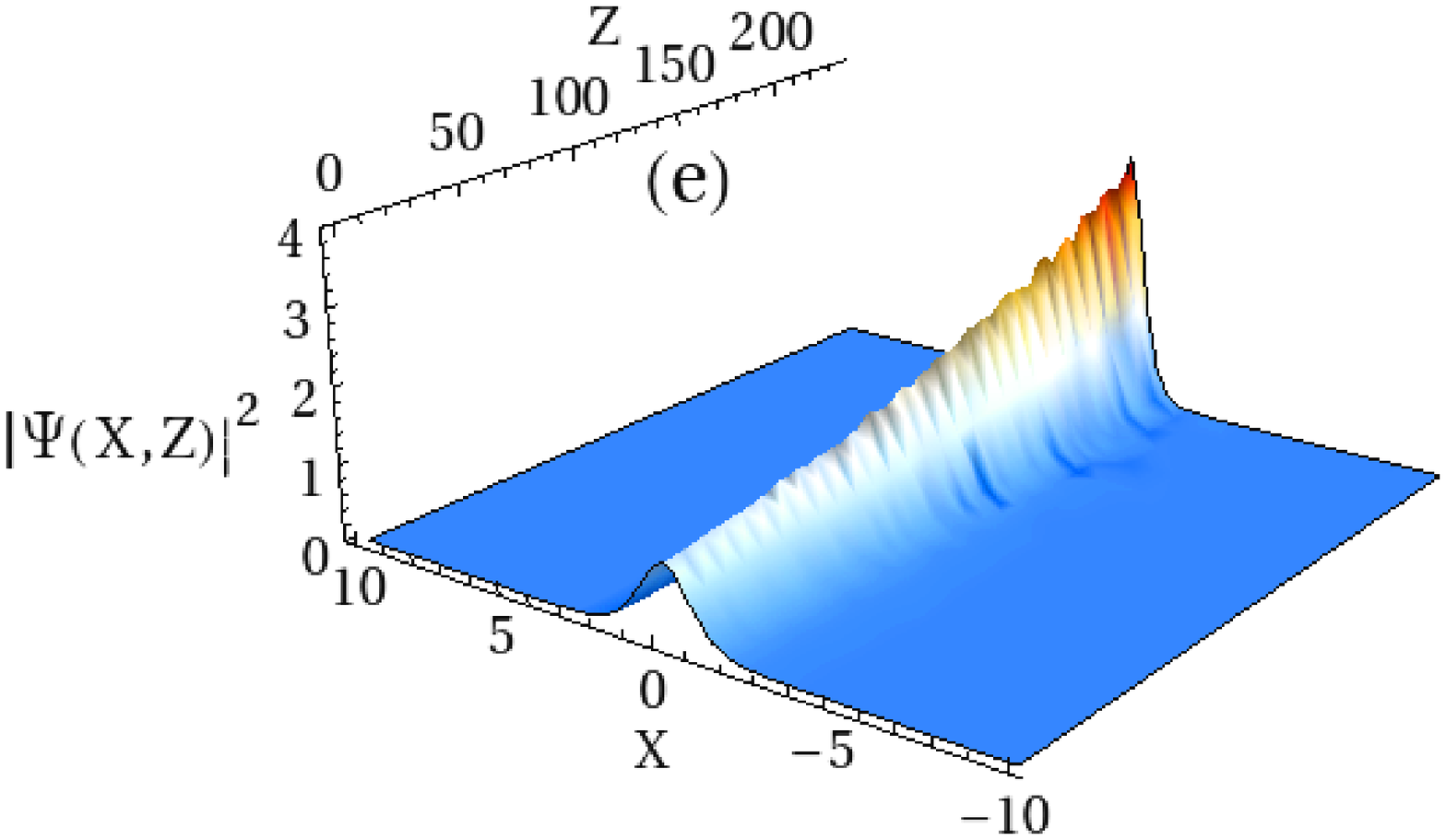} 
 \includegraphics[width=4.25 cm,height=4.5 cm]{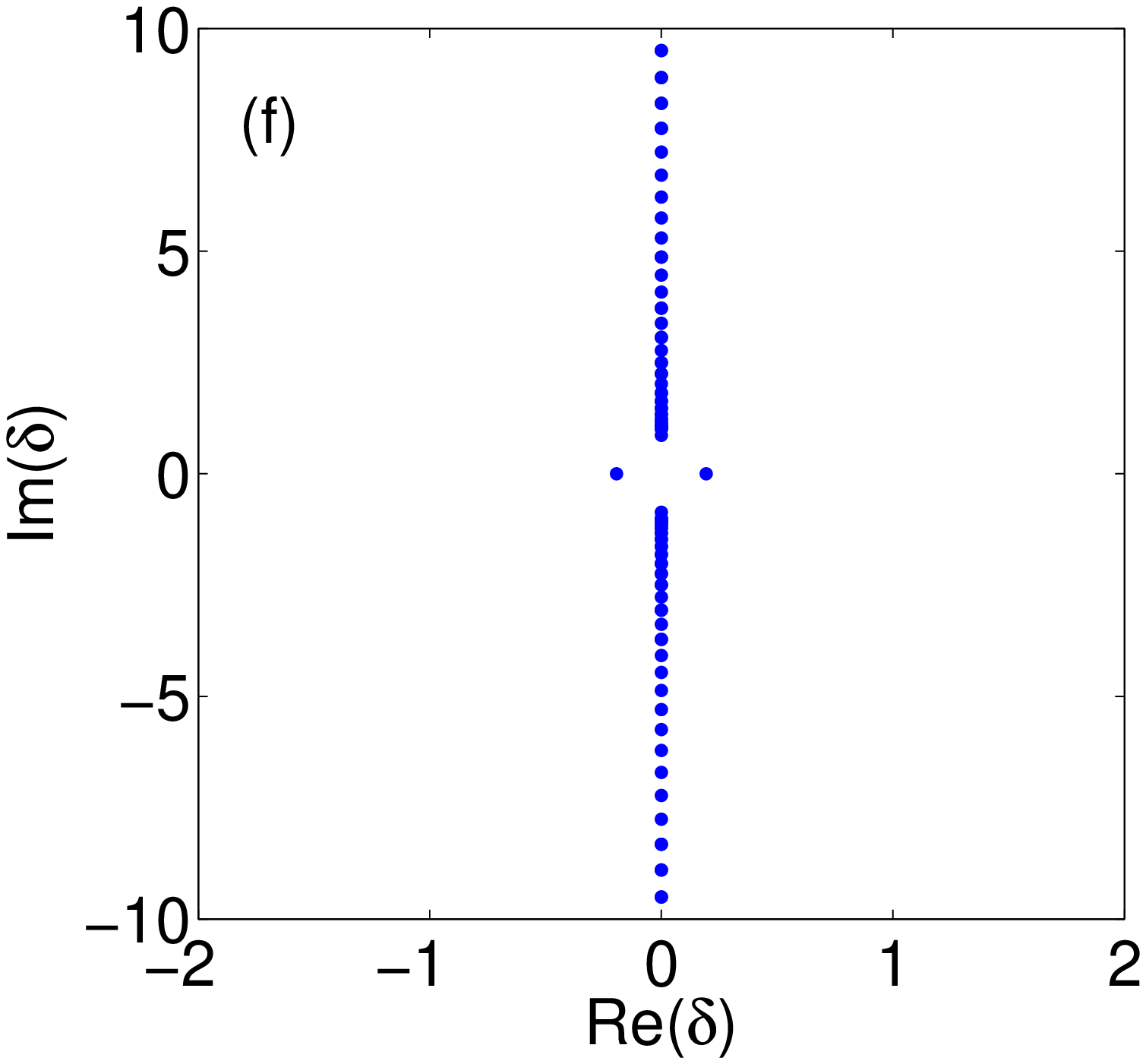}
  \caption{ (Color online) (a) Plot of the real and imaginary components of the stable localized modes $\phi(x)$ described by the 
  complex potential (\ref{e2}) for $k=3$, 
  (b) the stable evolution of the field intensity $|\Psi(x,z)|^2$, and (c) numerically obtained linear stability spectra. 
  All the figures have been plotted for 
 $k=3, V_0 =1, W_0 = .08, \sigma=1$ and $\beta =1$. (d), (e) (f) represent the plots of the same quantities as in (a), (b) and (c) but 
 for unstable case $W_0 =.9$.}\label{f3}
\end{figure}

In this case also we have the asymptotically vanishing gain and loss profile $W(x)$. Using the properties of the Hypergeometric function in equation 
(\ref{e4}) we have the localized modes corresponding to $k=3$ as $\phi(x) = \sqrt{2-V_0} \sech x ~ e^{\frac{i W_0}{10} [\tan^{-1} (\sinh x) + \sech x \tanh x ]}.$ 
For $V_0 =1$, we find that these localized modes are stable for $W_0 < .55$. Above this point the solitons become unstable after a long time of propagation.
In figure \ref{f3} (a),(b),(c), we have plotted such stable $\phi(x)$, evolution of the field intensity 
and the corresponding linear stability spectra. The same has been plotted in figures \ref{f3} (d),(e),(f) but for the unstable modes.\\

 Above analysis shows that the stability region increases with the increment of $k$. A probable explanation is that for larger values of $k$ 
the gain and loss profiles become 
more localized. Although we report here only the results of our analysis for some special cases of the competing parameter $k$, 
the stability analysis can be extended for the arbitrary positive real values of $k$. A detailed numerical investigation is required to find a minimum value of $k$ (if it exists) for which the nonlinear modes become completely stable.

\FloatBarrier

\subsection{Self-defocusing nonlinearity}
\begin{figure}[h]
  \includegraphics[width=4.25 cm, height=4.5 cm]{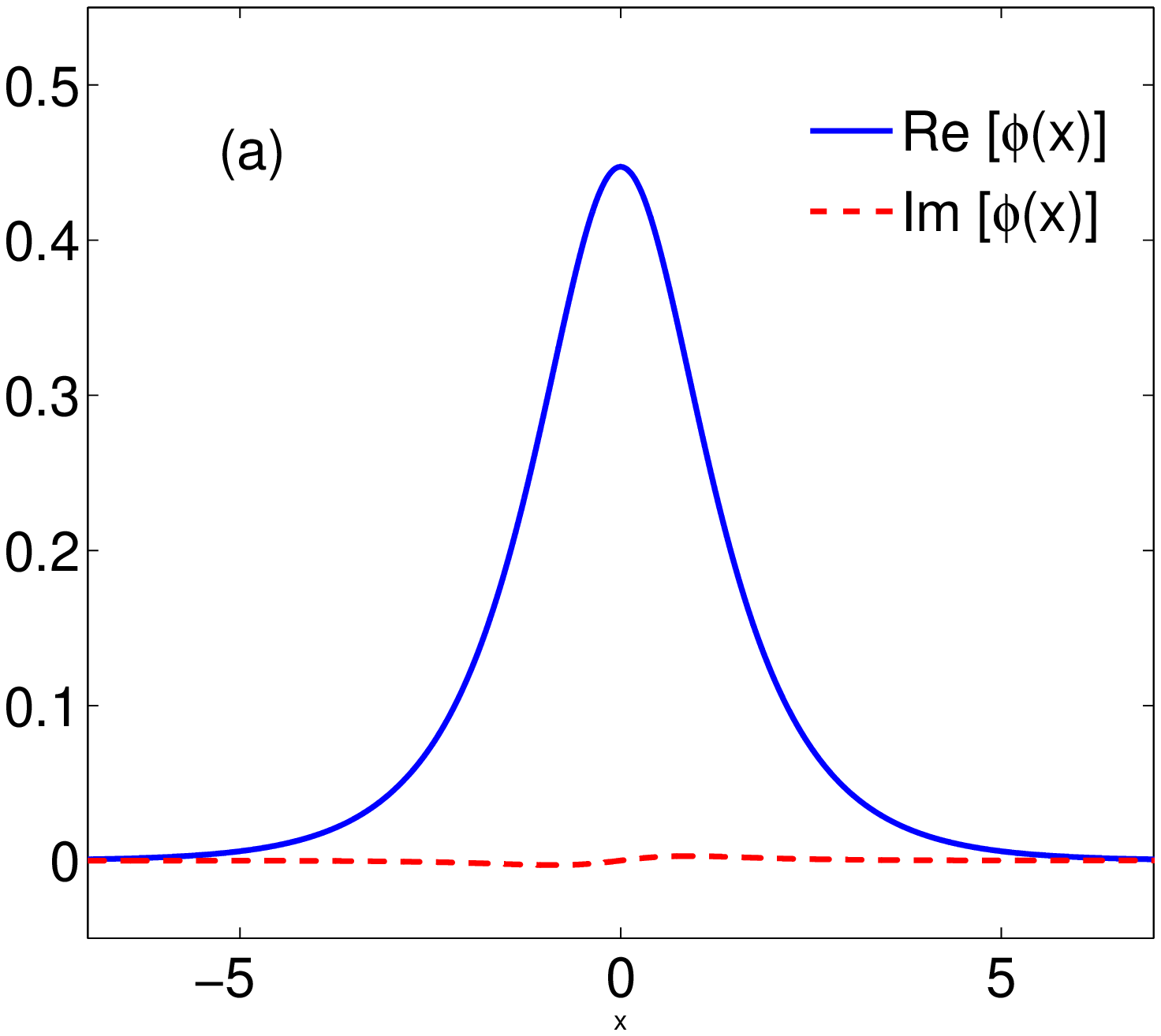} 
  \includegraphics[width=7.5 cm,height=4.5 cm]{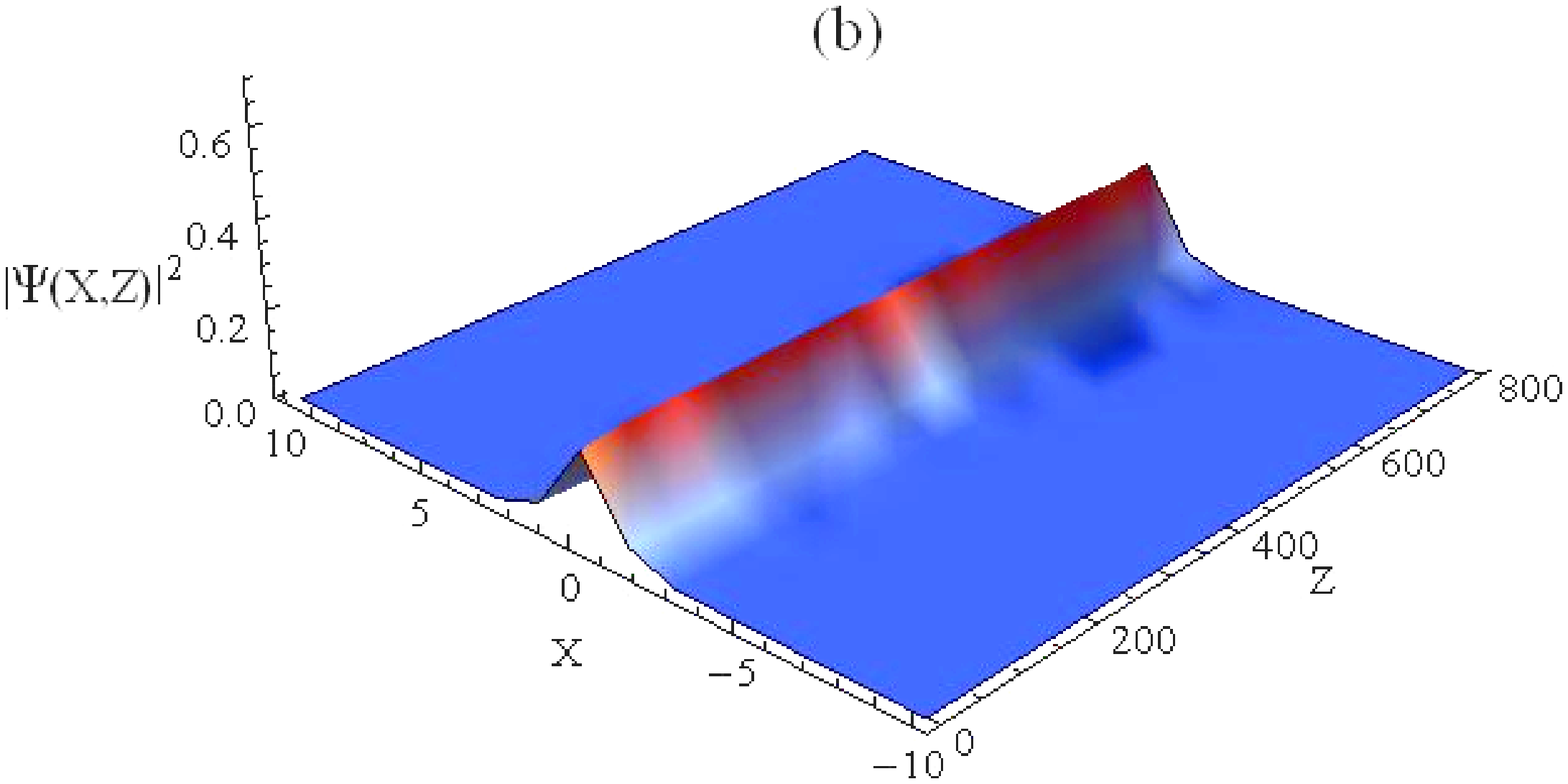} 
  \includegraphics[width=4. cm,height=4.5 cm]{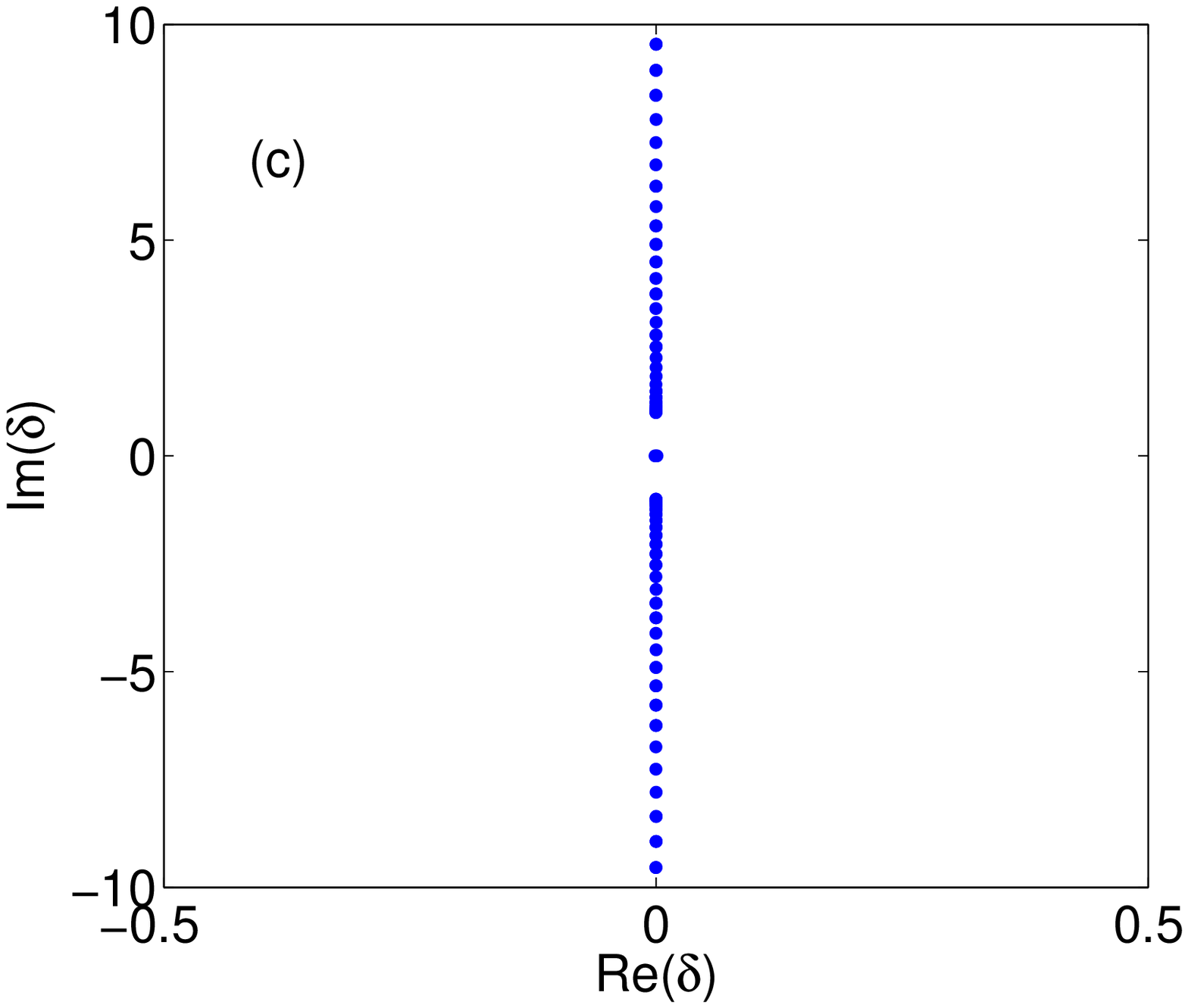}
  \caption{(Color online) Plot of (a) the real and imaginary components of the localized modes $\phi(x)$ for $k=2$ and in the prsence of defocusing nonlinearity; (b) the corresponding stable intensity evolution $|\Psi(x,z)|^2$, and (c) numerically computed linear stability spectra. 
  We have considered here $k=2, V_0 = 2.2, W_0 = 0.02, \sigma= -1$ and $\beta =1$. }\label{f7}
 \end{figure}

In the presence of self-defocusing nonlinearity the analytical expression of the localized modes is given by equation (\ref{e4}) with $\sigma=-1$. The total power $P = 2(V_0 -2)$ for such solutions remains positive for $V_0 > 2$. Linear stability analysis of the localized modes corresponding to self-defocusing nonlinearity reveals that these modes are always unstable for $k=0$. This case has also recently been reported by us in ref.\cite{MR13}. However, nonzero positive values of $k$ give rise to bright localized solution which are stable for some parameter values of $W_0$. Earlier in ref.\cite{Sh+11} it has been shown that stable bright soliton exist in $PT$-symmetric Scarf potential with self-defocusing Kerr nonlinerarity. Here we consider particular value $k=2$, for which real and imaginary parts of the potential and corresponding solution reduces to $V(x) = V_0 \sech^2 x + \frac{W_0^2}{16} \sech^4 x, ~ W(x) = W_0 \sech^2 x \tanh x,$ and $\phi(x)=  \sqrt{V_0-2} \sech x ~ e^{\frac{i W_0}{4} \tanh x}$, respectively. For $V_0=2.2$, the localized modes for this potential are found to be stable so long as  $0< W_0 <.1$. In figure \ref{f7},  we have shown such stable modes and corresponding intensity evolution for $W_0 = .02$.

 \FloatBarrier

\section{Localized modes in 2D}
\begin{figure}[h]
 \includegraphics[width=6.25 cm,height=4.5 cm]{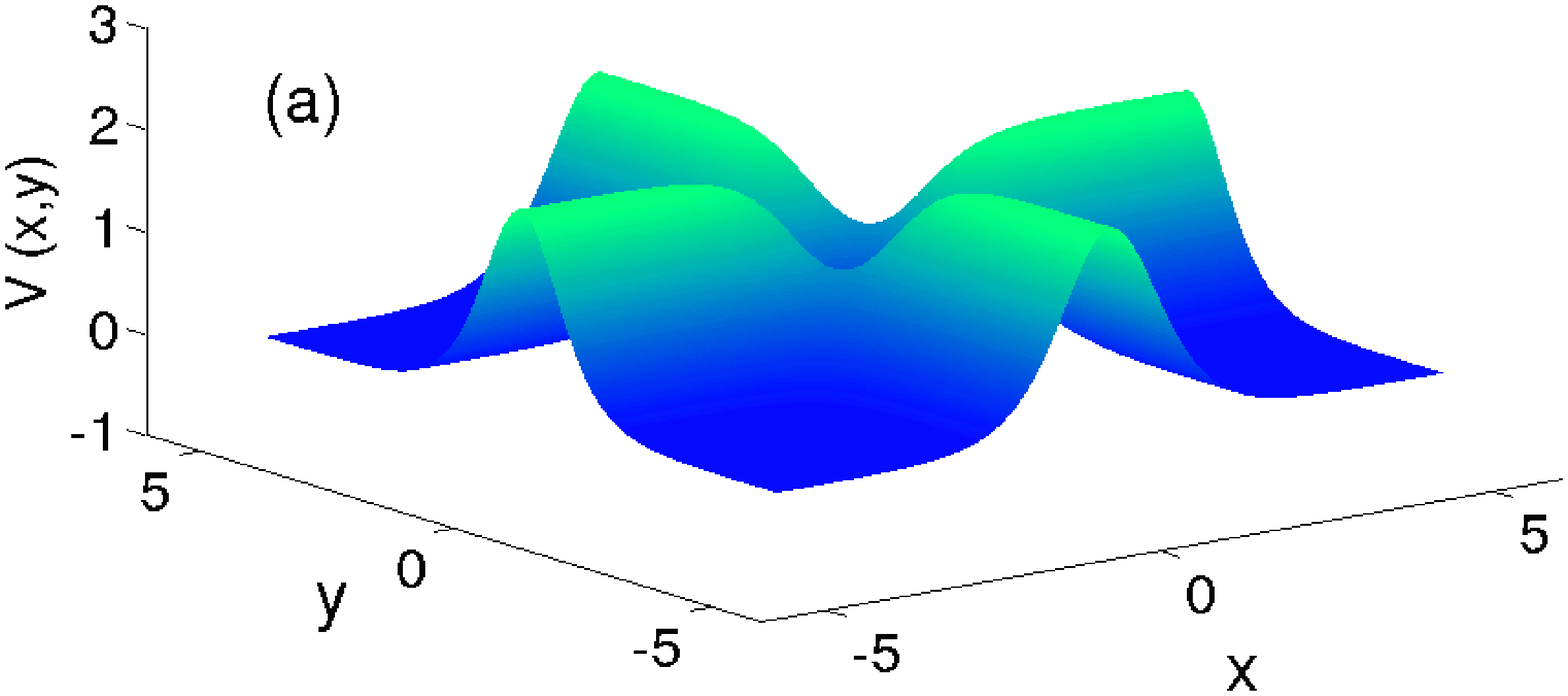} ~~~\includegraphics[width=7.25 cm,height=4.5 cm]{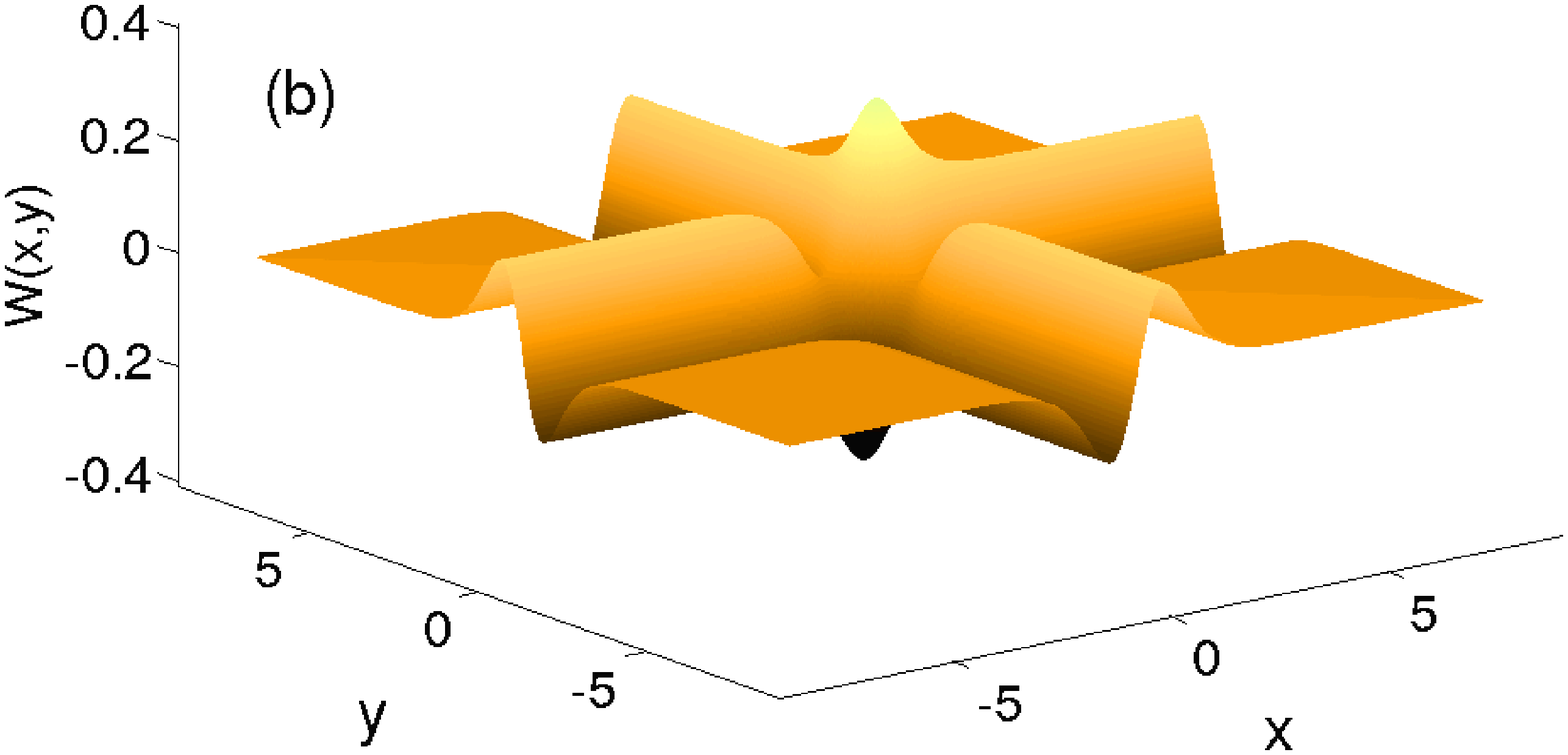}
 \includegraphics[width=7.25 cm,height=4.5 cm]{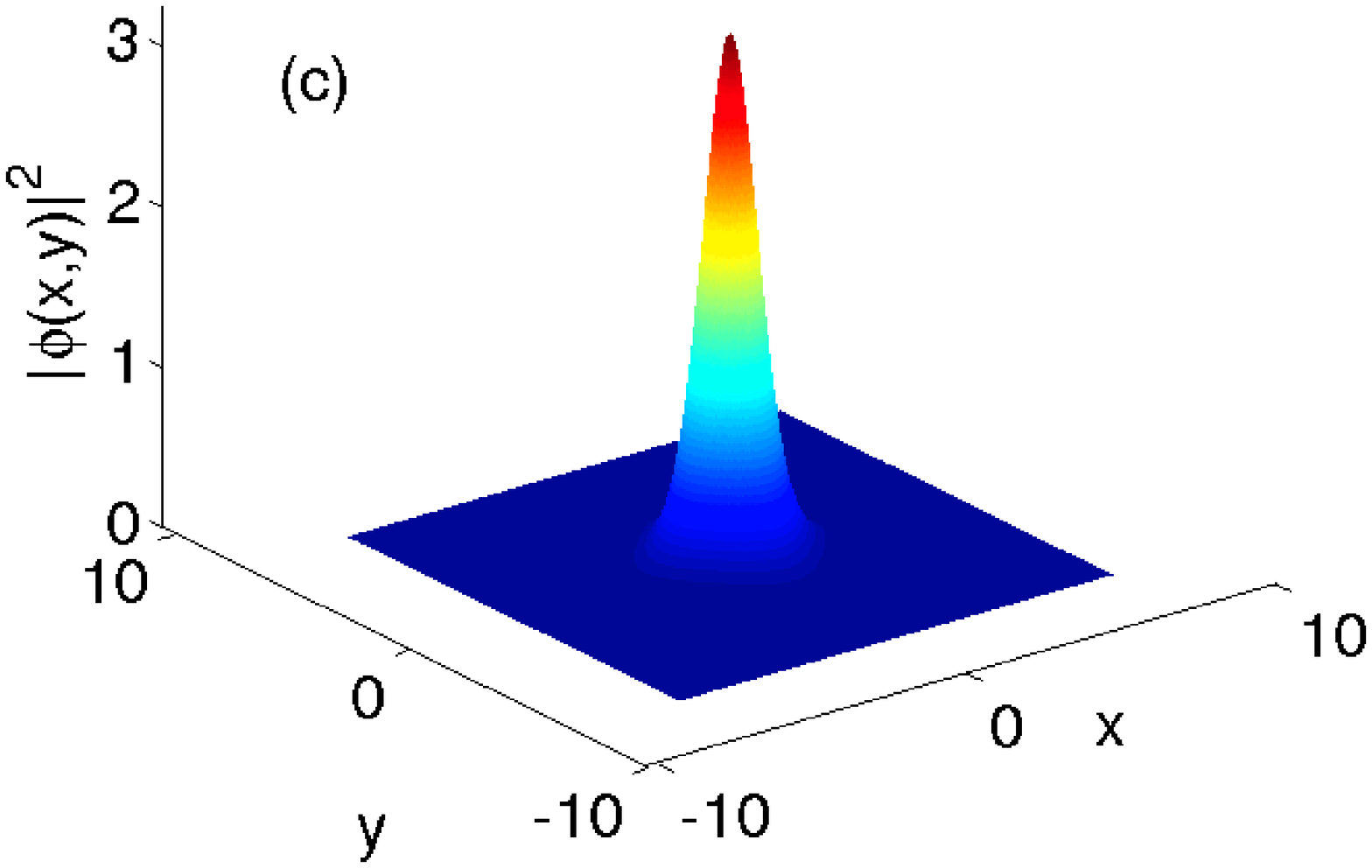} ~~~ \includegraphics[width=6.25 cm,height=4.5 cm]{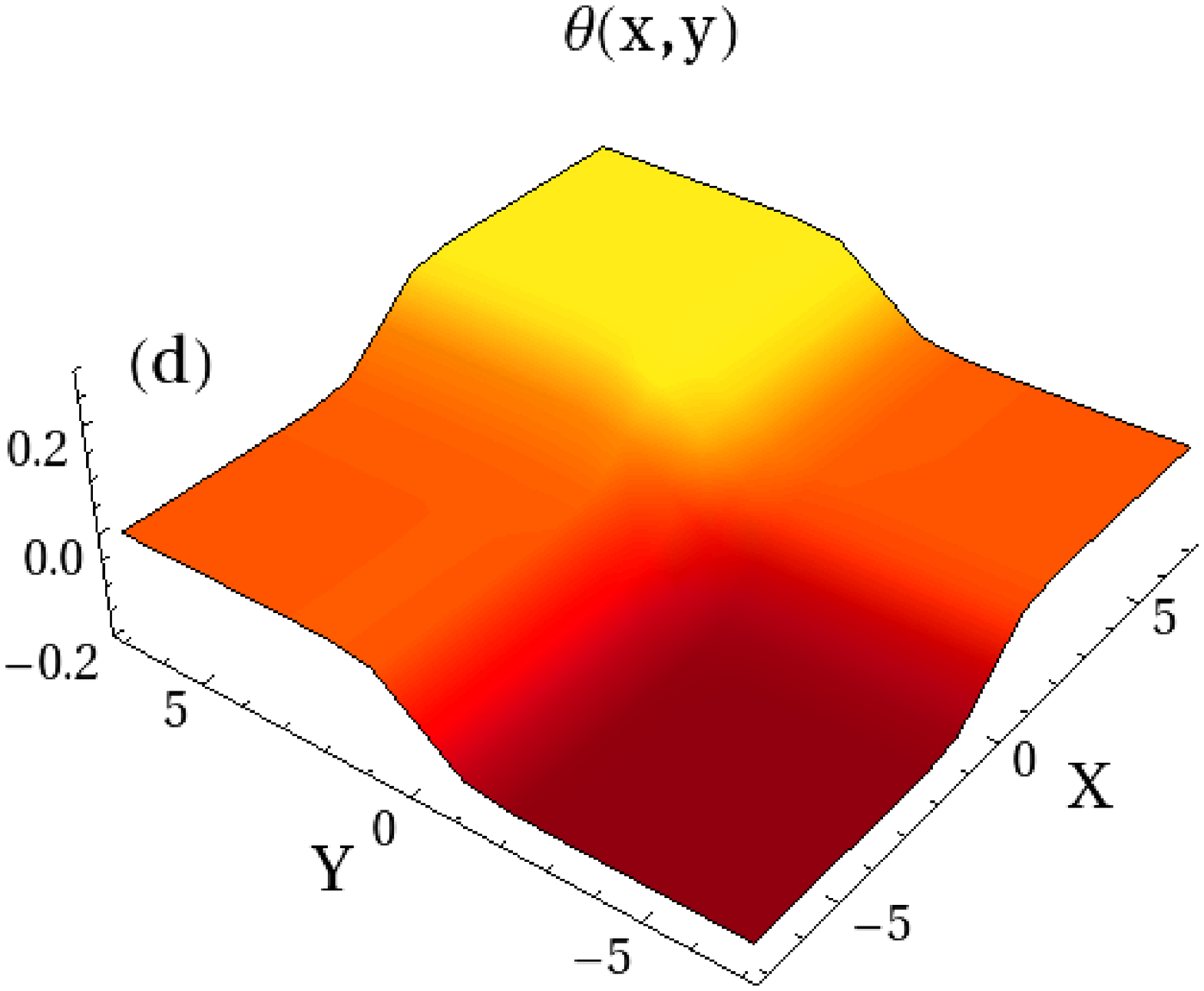}
 \caption{(Color online) (a), (b) Plot of the real and imaginary components of the 2D potential given in eq. (\ref{e11}).
 (c) Plot of the 2D localized modes $|\phi(x,y)|^2$ and (d) phase $\theta(x,y)$.
 In all these cases we have considered $k=4, V_0 = -1.8$, $W_0= 0.4, \beta = 2$ and $\sigma=1$.}\label{f4}
\end{figure}

The two-dimensional generalization of the potential (\ref{e2}) can be written as
\begin{equation}
 \begin{array}{ll}
  V(x,y) = 2 (\sech^2 x + \sech^2 y) + \frac{W_0^2}{(k+2)^2}(\sech^{2 k} x + \sech^{2 k} y)- (2-V_0) \sech^2 x ~ \sech^2 y\\
  W(x,y) =  W_0 (\sech^k x \tanh x + \sech^k y \tanh y).
 \end{array}\label{e11}
\end{equation}
which obey the $\mathcal{PT}$-symmetric condition $V(-x,-y) = V(x,y)$ and $W(-x,-y) = - W(x,y)$. For $k=4$, the real and imaginary parts of this potential are 
shown in figures 
\ref{f4}(a) and \ref{f4}(b), respectively. The stationary solution of the 2D self-focusing NLS equation
\begin{equation}
 i \frac{\partial \Psi}{\partial z} +  \nabla^2 \Psi + [V(x,y) + i W(x,y)] \Psi + |\Psi|^2 \Psi = 0, \label{e10}
\end{equation}
where $\nabla^2 \equiv \frac{\partial^2 }{\partial x^2} + \frac{\partial^2 }{\partial y^2}$ is the two-dimensional Laplacian, can be assumed 
in the form
\begin{equation}
 \Psi(x,y,z) = \phi(x,y) ~ e^{i \beta z + i \theta(x,y)}
\end{equation}
where the phase $\theta(x,y)$ and soliton $\phi(x,y)$ are real valued function and satisfy the following differential
equations
\begin{equation}\begin{array}{ll}
 \nabla^2 \phi - |\nabla \theta|^2 \phi + V(x,y) \phi + \phi^3 = \beta \phi, \\
 \phi\nabla^2 \theta + 2 \nabla \theta . \nabla \phi + W(x,y)\phi  = 0,
\end{array}\label{e22}
\end{equation}
respectively.
For the potential (\ref{e11}), the analytical solutions to equation (\ref{e22}) that satisfy $\phi \rightarrow 0$ as 
$(x,y) \rightarrow \pm \infty$ are obtained as
\begin{equation}
 \phi(x,y) =  \sqrt{2-V_0} \sech x \sech y,
\end{equation}
\begin{equation}
\theta(x,y)  =  \frac{W_0}{k+2} \left[\sinh x ~ _2F_1\left(\frac{1}{2},\frac{k+1}{2},\frac{3}{2},-\sinh^2 x\right) +
        \sinh y ~_2F_1\left(\frac{1}{2},\frac{k+1}{2},\frac{3}{2},-\sinh^2 y\right) \right],
\end{equation}
with the propagation constant $\beta = 2$. The plot of the two-dimensional soliton $|\phi(x,y)|^2$ 
and the phase $\theta(x,y)$ are shown in figure \ref{f4}(c) and \ref{f4}(d), respectively for $k=4$.
Here, the total power $P$ remains same as in the 1D case, whereas the 2D transverse power density across the beam is calculated as 
$\vec{S} = \frac{W_0}{k+2} (2-V_0) (\sech^{(k+2)} x, \sech^{(k+2)} y)$ and depends on competing parameter $k$. Also, this
suggests that the energy exchange from gain towards loss 
occurs if $W_0 >0$ and $V_0 < 2$. 
\FloatBarrier

\section{Summary}
To summarize, we have investigated the dynamics of nonlinear localized modes in optical media which is characterized by a class
of $\mathcal{PT}$-symmetric complex potentials with gain and loss components. Exact analytical expressions for these nonlinear modes have been 
obtained in the presence of self-focusing and self-defocusing Kerr nonlinearity. The linear stability analysis has been performed to investigate the 
stability of these localized modes. The results of this linear stability analysis have been verified by the direct numerical simulation of the 
NLS equation. Special emphasize has been given to the cases of asymptotically vanishing and non-vanishing gain/loss profiles. It is shown 
that the solitons associated with the former case are unstable (and thus are non physical) for all potential parameter values. In contrast to this, the solitons corresponding to the latter case propagate stably below a certain threshold value of $W_0$. For self-defocusing nonlinearity it is shown that stable bright soliton exist for some values of $k$. The localized modes in $2D$ and the transverse power flow density across
the beam are also determined. 

Finally, as has been observed in section IIB, the 
stability region increases as the value of the competing parameter $k$ is increased. A detailed numerical investigation is required to find minimum value of $k$ (if it exists) at which the nonlinear modes are stable for all values of $W_0$.  Moreover,  it should be mentioned here that the exact solution of the equation (\ref{e3}) has been obtained analytically for all values of $k$ but for a particular value of propagation constant $\beta$. It would be interesting to find generic solution valid for all propagation constant and to investigate the shape and stability of such solutions.

\end{document}